\documentclass[sigconf]{acmart}

\usepackage{enumitem}
\usepackage[english]{babel}
\usepackage{hyperref}

\addto\extrasenglish{%

}
\usepackage{multirow}
\usepackage[capitalize]{cleveref}

\crefformat{table}{(#2Table~#1#3)}
\crefformat{figure}{(#2Figure~#1#3)}
\usepackage{graphicx}
\usepackage{balance}
\usepackage{subcaption}
\usepackage{amsmath}
\usepackage{lineno}
\usepackage{color}
\usepackage{setspace}
\usepackage{tikz}
\usepackage{multirow}
\usepackage{rotating}
\usetikzlibrary{shapes, backgrounds}
\usepackage{xcolor}
\usepackage{booktabs}

\newcommand{\revision}[1]{\textcolor{black}{#1}} 
\AtBeginDocument{%
  }

\setcopyright{acmlicensed}
\copyrightyear{2026}
\acmYear{2026}
\acmDOI{XXXXXXX.XXXXXXX}
\acmConference[CHIWORK '26]{CHIWORK '26: Proceedings of the 5th Annual Symposium on Human-Computer Interaction for Work}{June 22--25, 2026}{Linz, Austria}

\acmISBN{978-1-4503-XXXX-X/2018/06}

\begin{document}

\title{Confidence Without Competence in AI-Assisted Knowledge Work}

\author{Elena Eleftheriou}
\orcid{0009-0007-7632-4262}
\affiliation{%
  \institution{University of Cyprus}
  \city{Nicosia}
  \country{Cyprus}}
\email{eleftheriou.elena@ucy.ac.cy}

\author{George Pallis}
\orcid{0000-0003-1815-5468}
\affiliation{%
  \institution{University of Cyprus}
  \city{Nicosia}
  \country{Cyprus}}
\email{pallis.george@ucy.ac.cy}

\author{Marios Constantinides}
\orcid{0000-0003-1454-0641}
\affiliation{
  \institution{CYENS Centre of Excellence}
  \city{Nicosia}
  \country{Cyprus}}
\authornote{Also affiliated with University of Cyprus, Cyprus and University College London, UK.}
\email{marios.constantinides@cyens.org.cy}

\renewcommand{\shortauthors}{Eleftheriou et al.}

\begin{abstract}

Large Language Models (LLMs) are widely used by students, yet their tendency to provide fast and complete answers may discourage reflection and foster overconfidence. We examined how alternative LLM interaction designs support deeper thinking without excessively increasing cognitive burden. We conducted a two-phase mixed-methods study. In Phase 1, interviews with 16 Gen Z students informed the design of Deep3, a web-based system with three interaction modes: \emph{a)} future-self explanations, \emph{b)} contrastive learning, and \emph{c)} guided hints. In Phase 2, we evaluated Deep3 with 85 participants across two learning tasks. We found that a standard single-agent baseline produced high perceived understanding despite the lowest objective learning. In contrast, future-self explanations imposed higher cognitive workload yet yielded the closest alignment between perceived and actual understanding, while guided hints achieved the largest learning gains without a proportional increase in frustration. These findings show that effort, confidence, and learning systematically diverge in LLM-supported work.


\end{abstract}



\begin{teaserfigure}
    \centering
    \includegraphics[width=1\linewidth]{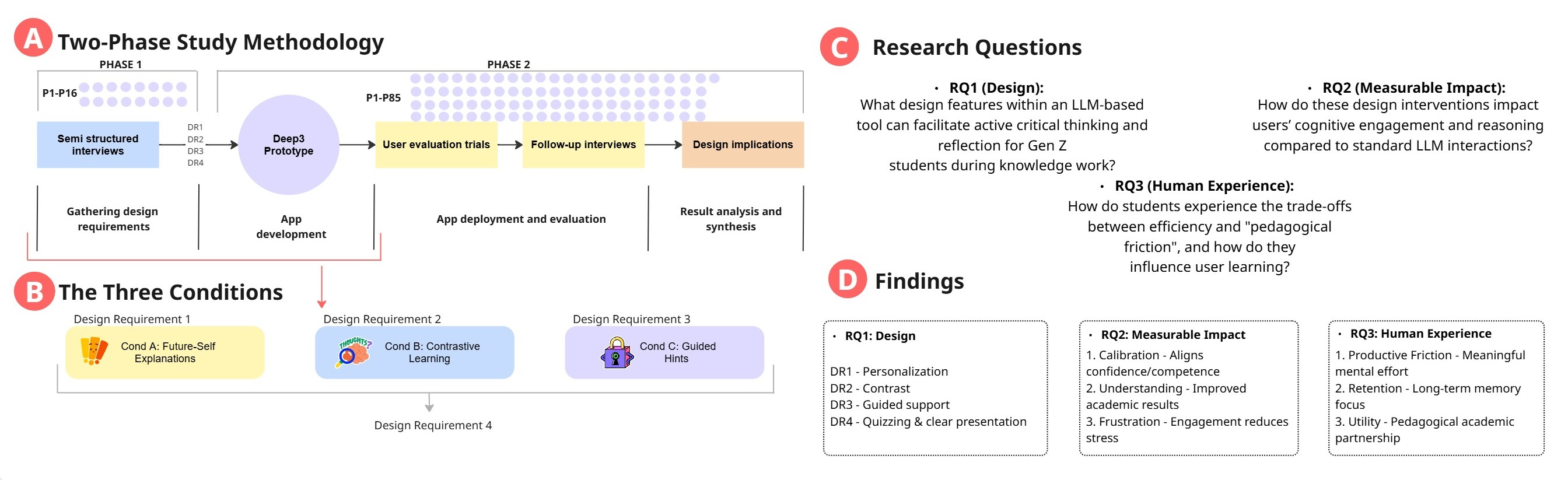}
    \caption{Overview of our study and findings, which illustrate (A) the two-phase study methodology; (B) the three conditions developed from the design requirements gathered in Phase 1; (C) our research questions; and (D) the resulting findings.}
    \label{fig:placeholder}
\end{teaserfigure}


\maketitle

\section{Introduction}
\label{sec:introduction}

The rapid integration of LLMs into higher education has fundamentally changed the academic environment for Generation Z (Gen Z) and emerging young professionals. As a ``digital-first'' generation, these students have quickly adopted tools like ChatGPT to speed up information retrieval, note-taking, and problem-solving, treating generative AI (GenAI) as ``the new calculator'' for the modern era~\cite{simkute2025new}. While recent studies show that nearly all Gen Z students utilize these tools for tasks ranging from note-taking to test preparation~\cite{amoah2025chatgpt,olander2025}, this adoption has created a generational divide. Students embrace the efficiency of AI, while educators express growing concern over the ethical and pedagogical shifts this  ``calculator'' mentality introduces to the classroom~\cite{chan2023ai}.

Despite the undeniable gains in productivity, the widespread reliance on LLMs risks a major loss of meaningful cognitive growth. Current research suggests that when AI handles the complex cognitive demands of knowledge work, it can lead to ``displaced thinking'', which is a phenomenon where users experience lower brain engagement and a reduced capacity for critical evaluation~\cite{georgiou2025chatgpt,kosmyna2025your}. As~\citet{sarkar2024copilot} argue in their exploration of how to stop AI from killing critical thinking, there is a danger that AI tools act as ``autopilots'' that bypass the necessary friction required for deeper learning. When users prioritize speed over understanding, they often fall into a ``dependency trap'', a state in which high confidence in AI results correlates with reduced critical reasoning~\cite{gerlich2025ai,lee2025impact}. This “dependency trap” is reinforced by an illusion of competence, where reliance on seamless AI outputs leads users to stop thinking for themselves, and fail to notice errors or missing logic.~\cite{LiraEtAl2026GenZGenAI}.

To address this gap, our work investigates how the integration of LLMs can be structured to support the cognitive development of Gen Z students rather than bypassing it. Instead of proposing a total ban on AI, our focus is on investigating how ``pedagogical friction'' can be integrated into existing LLMs to accommodate the demands of modern higher education. We position our work as an exploratory study that surfaces the trade-offs between different agents and student engagement. We aim to surface lived user experiences and provide practical design implications for maintaining critical reasoning in an era of automated knowledge work.

To achieve this, we carried out a two-phase mixed-methods study. In Phase 1, we conducted semi-structured interviews with 16 Gen Z students \revision{studying at universities across Europe}, during which we explored how they used LLMs during their studies, and how could a tool help them engage more actively in thinking and learning. Their insights informed the design of Deep3, a web-based system offering deliberate, critical and analytical reflection across three different agents~\cref{fig:placeholder}. Deep3 was designed to accommodate various modes of academic engagement, allowing users to select a specific interaction style based on their current study goals with the AI assistant providing tailored suggestions for each mode. In Phase 2, we evaluated the Deep3 application through a between-subjects experiment with 85 participants, \revision{at European universities}. We utilized a mixed-methods approach, combining interaction log files, pre- and post-task questionnaires, and semi-structured follow-up interviews, to examine how our different agents impact student understanding and cognitive load. Specifically, we address three research questions (RQs):

\begin{itemize}
    \item[\textbf{RQ$_{1}$:}] What design features within an LLM-based tool can facilitate active critical thinking and reflection for Gen Z students during their academic work?
    
    \item[\textbf{RQ$_{2}$:}] How do these design interventions impact users’ cognitive engagement and reasoning compared to standard LLM interactions?
    
    \item[\textbf{RQ$_{3}$:}] How do students navigate the tension between speed and this new friction, and how does it influence user learning?
\end{itemize}

We designed two tasks that required active cognitive engagement and could benefit from pedagogical friction. In the first task, the user engaged with the system to clarify an academic concept until they could demonstrate a sufficient level of understanding. In the second, the user had to solve a logic-based puzzle with the help of the agent, requiring them to compare different steps and ideas to find the correct solution. Together, these tasks model interaction scenarios that align with documented LLM usage patterns among university
students~\cite{paustian2024students,ravvselj2025higher}.

In answering our RQs, we made two main contributions:

\begin{enumerate}[label={(\arabic*)}]
    \itemsep0.5em 
    \item  Through semi-structured interviews with Gen Z students (\autoref{sec:formative-study}), we identified four design requirements for an LLM-based tool capable of fostering critical thinking. With these requirements at hand, we designed and deployed a web-based system called \textit{Deep 3} (\S\ref{sec:App}).
    
    \item We conducted a between-subjects evaluation comparing four agent configurations (\S\ref{sec:user-evaluation}) to understand how these designs impact student reasoning, cognitive load, and understanding compared to a standard LLM. We found that while a standard LLM satisfies a ``speed addiction'' with a misalignment between confidence and competence, an LLM with pedagogical friction can close this gap, enhancing actual understanding without necessarily increasing user frustration (\S\ref{sec:results}).
\end{enumerate}

In light of these results, we conclude with empirical and design insights on how pedagogical friction shapes the cognitive and behavioral dimensions of student interactions with distinct LLM agent designs (\S\ref{sec:discussion}). \revision{While our findings are grounded in higher education, they offer a blueprint for designing AI tools in broader work contexts, specifically for professional sectors with a fast-paced environment that require active human-in-the-loop reasoning.}

\section{Related Work}
\label{sec:related-work}
We reviewed the key research areas that inform our work and organized them into three areas: i) Gen Z and LLMs in education (\S\ref{subsec:related1}); ii) the cognitive impact of LLM use (\S\ref{subsec:related2}); and iii) technologies and techniques for reflection (\S\ref{subsec:related3}).

\subsection{Gen Z and LLMs in Education} \label{subsec:related1}
Gen Z, born between 1995 and 2012, is the first generation to grow up with constant access to digital technology and social media, which gives them a digital-first mindset, as~\citet{puiu2017generation} describes. As a result, Gen Z students often expect that educational institutions will provide up-to-date technological tools and resources to support their learning. Given this, it is becoming widely accepted that  GenAI technologies, particularly LLMs such as ChatGPT, have the potential to transform education by improving teaching and learning~\cite{kasneci2023chatgpt}.

According to recent studies, Gen Z students are using AI tools extensively for their academic work. For example, a large-scale survey reported that 97\% of Gen Z students use them, with 66\% to study, 56\% to prepare for tests and 46\% to take notes~\cite{olander2025}. The 2025 HEPI Student Survey indicates that student engagement with GenAI for assessments has jumped from 53\% last year to 88\% this year, with a primary focus on concept explanation and study support~\cite{freeman2025student}.



However, this integration is not without friction. While research shows that Gen Z students are generally optimistic about the benefits of LLMs, expressing strong intentions to integrate these tools into their learning practices~\cite{chan2023ai}. At the same time, Gen X and Millennial teachers tend to approach these tools with greater caution, raising concerns about ethical and pedagogical implications~\cite{chan2023ai}. This tension is most noticeable in how people use their thinking skills. Teachers are increasingly worried about what they call ``metacognitive laziness'', meaning that students are not actively thinking~\cite{nosta2025shadow}. This generational divide clearly shows the tension that LLMs introduce regarding cognitive use.

\subsection{The Cognitive Impact of LLM Use} \label{subsec:related2}
The widespread adoption of LLM tools raises questions about how they may affect cognitive processes, critical thinking and learning outcomes. Recent studies suggest that reliance on LLMs for gathering information and problem-solving can both support and hinder learning. From one perspective, LLMs can improve a student’s understanding by providing explanations, but they may also increase reliance on incorrect answers~\cite{kim2025fostering}.

Over-reliance may also reduce student's critical reasoning. For example,~\citet{kosmyna2025your} found that participants, who used LLMs, showed lower brain engagement and less critical evaluation of their work compared to those who completed tasks without it. These findings were validated in 2025 by the MIT Media Lab, which used EEG monitoring to show that students enter a ``neural standby mode'' when using AI, leading to poor memory retention~\cite{Fan2024MetacognitiveLaziness}. 

Similarly, another study revealed that higher confidence in GenAI was associated with less critical thinking~\cite{lee2025impact}.~\citet{lee2025impact} expanded on this, finding a direct negative correlation between trust in AI and the performance of critical verification steps.~\citet{gerlich2025ai} also reported that frequent use of LLM tools, especially for younger users, was linked to weaker critical thinking skills.

~\citet{georgiou2025chatgpt} observed that ChatGPT use resulted in significantly lower cognitive engagement scores compared to participants who did not use it. Lastly,~\citet{rogers2025worth} highlighted the importance of effort in learning, suggesting that requiring students to engage more actively with LLM tools could improve learning outcomes. This is supported by~\citet{Hong2025CognitiveOffload}, who found that ``Cognitive Offload Instruction'', where AI is restricted to low-level tasks, can actually preserve and enhance high-level critical thinking. Together, these findings highlight the risks of overreliance on GenAI. 

\subsection{Technologies and Techniques for Reflection} \label{subsec:related3}
Research has consistently shown that self-explanation supports learning by enabling deeper understanding and problem-solving skills.~\citet{chi1989self} found that students who explained ideas to themselves while studying performed better on tests because it helped them notice gaps in their knowledge and connect new concepts. Similarly,~\citet{mceldoon2013self} showed that prompting students to self-explain during math problems led to better conceptual understanding than simply giving them more practice. Recent large-scale research in digital learning environments (DLEs) further validates these findings, demonstrating that this strategy consistently improves academic performance~\citet{tan2025enhancing}. Overall, self-explanation supports critical thinking and deeper learning.

Recent HCI research has explored ways to keep users actively engaged and reflective rather than putting knowledge work on ``autopilot'' ~\cite{sarkar2024copilot}. One strategy is to turn the LLM into a critic by having it reflect on its own outputs.~\citet{sarkar2024copilot} built a spreadsheet tool where the LLM suggests an action and then provides a short ``provocation'', that is a short critique highlighting possible issues or alternatives. This idea led users to pause, reflect, and think critically before accepting LLM suggestions~\cite{drosos2025makes}. Similarly,~\citet{buccinca2025contrastive} showed that contrastive explanations, those that compare the LLM’s chosen answer to a predicted human response, can improve decision-making without reducing accuracy. Moreover, the "Judge-Questioner" framework employs dual LLMs to generate critical questions that challenge a student's unsupported claims, significantly increasing engagement with argumentative texts~\cite{favero2025ellis}.

Another approach is to have LLMs ask questions instead of giving direct answers.~\citet{danry2023don} show that when LLM explanations are framed as questions, people become better at spotting flaws in reasoning. In this way, LLMs act more like a ``critical thinking coach'' than just an ``information provider''. Similarly,~\citet{park2023thinking} describe ``Thinking Assistants'' as interactive agents that use LLM dialogue to generate guided and domain-specific questions. Their experiments found that participants using these assistants reflected more on their thinking compared to those using answer-focused agents. This question-driven approach can improve learning, though studies show that questions alone may increase engagement without additional learning gains~\cite{blasco2024ai}.

Several LLM tutoring systems are also designed to require reflection. For instance, Iris Tutor~\cite{bassner2024iris} for programming tasks gives hints and counter-questions instead of full solutions. It uses chain-of-thought prompting with awareness of the student’s code to adapt its advice. Students reported that Iris was effective and boosted their confidence. Similarly, TutorLLM~\cite{li2025tutorllm} adapts its explanations to each student’s progress by tracking what they know. This approach helps students better understand the material, feel more satisfied with the learning experience, and perform better on quizzes. The SocratiQ~\cite{jabbour2025socratiq} system also uses a Socratic framework, creating personalized learning paths through interactive questioning. In all these systems, LLMs guide students to solve problems and explain their steps, rather than passively consuming answers.
\smallskip

\noindent\textbf{Research Gaps.} Despite growing interest in LLM-supported learning and reflection, several gaps remain in existing literature. Firstly, while self-explanation has been widely shown to enhance understanding and critical thinking, there is no recent HCI research exploring how this strategy might function when integrated with AI tools. Secondly, previous work has explored methods for prompting users to reflect through critique or contrast~\cite{buccinca2025contrastive}. However, these approaches typically rely on pre-generated statements, \revision{and not dynamic ones.}
Finally, prior work on question-driven or Socratic LLMs has shown that asking users questions can increase engagement, but these systems often lack mechanisms for additional learning gains 
~\cite{blasco2024ai, bassner2024iris}. Our study addresses these gaps by examining how different forms of LLM interaction can support active reflection and critical thinking during knowledge work.

\section{Author Positionality Statement}
We recognize that our positionality shaped the study's framing, design, and interpretation~\cite{singh2025exploring, havens2020situated}. Our team is comprised of researchers in Human–Computer Interaction with backgrounds spanning human-centered AI, data science, learning technologies, and empirical studies of digital work practices. The first author is a Generation Z student who regularly uses Large Language Models as part of their academic work. Their lived experience with AI-assisted studying informed the initial motivation for this research and the framing of overconfidence, effort, and reliance on AI tools. The remaining authors are researchers in Human–Computer Interaction with backgrounds in human-centered AI, learning technologies, and empirical studies of digital work practices. While not members of Generation Z, they engage extensively with Gen Z students through teaching, supervision, and research, all of which shape a complementary perspective grounded in pedagogy, design, and critical reflection.

These differing perspectives influenced the study design and interpretation. The first author's experiences helped surface everyday practices and tensions around AI use, while the broader research team emphasized methodological rigor and balanced interpretation. To mitigate potential biases, the study combined qualitative accounts with objective learning measures, foregrounds participant perspectives, and reports trade-offs across interaction designs rather than advocating a single optimal solution.

\section{Formative Study: Design Requirements}
\label{sec:formative-study}

\begin{table*}[t!]
    \centering
    \caption{Participant demographics, including age, gender, education, field of study, and frequency of LLM interaction.}
    \label{tab:demographics}
    \scalebox{1}{
    \begin{tabular}{lllllll}
    \toprule
    \textbf{ID} & \textbf{Age} & \textbf{Gender} & \textbf{Education} & \textbf{Field of Study} & \textbf{LLM Interaction Frequency} \\ 
    \midrule
    1  & 22 & Female & Bachelor's degree (in progress) & Computer Science & Often (a few times a week) \\
    2  & 23 & Female & Master's degree (in progress) & Philology & Sometimes (a few times a month) \\
    3  & 23 & Female & Master's degree (completed) & Philology & Daily \\
    4  & 22 & Male & Bachelor's degree (completed) & Computer Science & Daily \\
        5  & 21 & Male & Bachelor's degree (in progress) & Computer Science & Often (a few times a week) \\
    6  & 21 & Male & Bachelor's degree (in progress) & Computer Science & Daily \\
    7  & 21 & Female & Bachelor's Degree (in progress) & Civil Engineering & Daily \\
    8  & 21 & Female & Bachelor's degree (in progress) & Civil Engineering & Daily \\
    9  & 22 & Male & Bachelor's degree (in progress) & Computer Science & Daily \\
    10 & 21 & Female & Bachelor's degree (completed) & Chemical Engineering & Daily \\
    11 & 22 & Male & Bachelor's degree (in progress) & Chemical Engineering & Sometimes (a few times a month) \\
    12 & 22 & Male & Bachelor's degree (in progress) & Mathematics & Sometimes (a few times a month) \\
    13 & 22 & Female & Bachelor's degree (in progress) & Computer Engineering & Often (a few times a week) \\
    14 & 21 & Female & Bachelor's degree (completed) & Economics & Sometimes (a few times a month) \\
    15 & 23 & Female & Bachelor's degree (completed) & Electrical Engineering & Daily \\
    16 & 22 & Male & Bachelor's degree (in progress) & Business & Daily \\
    \bottomrule
    \end{tabular}%
    }
\end{table*}

To identify the requirements for our web application, we followed a two-step method that combined insights from established literature with findings from a targeted formative study. This allowed us to first identify broad patterns in how Gen Z students use LLMs and then explore how a tool could specifically help them engage more actively in thinking and learning.

\subsection{Step 1: Insights from Literature on Gen Z and LLMs}

To establish a baseline for Gen Z’s interaction with AI, we first reviewed recent large-scale surveys and meta-analyses, which indicated increasing integration of AI into everyday learning activities. Data from a survey~\cite{ScholarshipOwl2025GenZAI} of over 12,000 students confirms a near-universal adoption rate (97\%). The majority of students utilize AI as a ``Strategic Academic Assistant'' for test preparation (56\%) and searching for scholarly literature (55\%). However, while this usage significantly boosts immediate academic performance ($g = 0.867$), a meta-analysis by~\citet{wang2025effect} indicates a much lower impact on critical thinking ($g = 0.457$), suggesting that students often stop at memorization instead of developing a deeper understanding. This ``thinking gap'' is further refined by the affordance-based research of~\citet{lee2025examining}. They found that Gen Z particularly uses AI to generate quick ideas and overcome the psychological barrier of starting with a blank page. 

\subsection{Step 2: Formative Study}
While Step 1 provided the ``what'' and ``how many'', we conducted semi-structured interviews with Gen Z students to explore how a tool could help them engage more actively in thinking and learning. The study helped us identify specific design requirements. \revision{The formative study took place in September 2025.}

\subsubsection{Participants} 
We recruited 16 Gen Z students (P1-P16) through university connections using convenience sampling. \revision{Participation was entirely voluntary, and all individuals provided informed consent prior to the study.} The study was approved by the \revision{Cyprus National Bioethics Committee}. All participants were enrolled in a bachelor or master program and represented a diverse range of academic disciplines, including Philology, Economics, Computer Science, Electrical Engineering, Civil Engineering and Chemical Engineering. They were aged 21-23 years, with 9 female and 7 male students. Participants reported moderate to high interaction frequency with AI models, rating their usage between 3 to 5 on a 5-point scale. All participants' demographics are presented in Table~\ref{tab:demographics}. 

\subsubsection{Procedure}
Each participant completed a demographic survey prior to the semi-structured interviews. Interviews lasted 20-30 minutes and took place either online or in person, depending on participant preference. The interview protocol was informed by prior literature on LLM-assisted learning and reflective tool use, particularly studies exploring how LLMs can support critical thinking and engagement~\cite{blasco2024ai,danry2023don,drosos2025makes,park2023thinking}. \revision{The protocol can be found in Appendix~\ref{app:interview_questions}.}

\subsubsection{Data Collection and Analysis} Interviews were recorded, transcribed using \href{https://www.rev.com}{rev.com}, and reviewed for accuracy. We used thematic analysis for its flexibility in exploring user perspectives~\cite{braun2006using}. We began with repeated reading and annotation to develop codes (e.g., critical reflection, engagement strategies, interaction style with LLMs). The codes were iteratively grouped into themes through discussion. By synthesizing them with large-scale trends from literature, we identified four design requirements (DRs), three of which were focused on functionality and one on presentation. 

\subsubsection{Design Requirements} 

Participants emphasized the importance of a learning tool that adapts to individual knowledge levels and preferences. Users described wanting an experience that feels like \textit{``asking a question to yourself but to a more professional you''} (P16), highlighting the need for the system to adjust explanations to each learner’s needs. They also expressed that the tool should communicate in ways that align with each person’s thinking and preferences, making answers more understandable and natural, as P12 described, \textit{``This will be simpler and more easy to use because it will be like I'm talking to myself but myself that knows a lot more''}. These insights inform our first design requirement, which focuses on \revision{adaptive personalization}, ensuring that questions and answers are tracked in a clear and simple way. This design requirement is supported by prior research on adaptive learning environments. Adaptive systems that adjust feedback and content based on learners’ knowledge levels have been shown to improve engagement and learning performance~\cite{alshammari2019effective}. As~\citet{nunez2026one} argue, these systems succeed by bridging the gap between plain technology and teaching methods, allowing for a ``customized teaching'' approach. This approach directly reflects the ``Content Curation'' capability identified by~\citet{lee2025examining}, where Gen Z students prefer AI interactions that resemble personalized tutoring rather than static search results.

\begin{quote}
    \textbf{DR1: \revision{Adaptive Personalization}} The system should adjust to the user’s knowledge level, and tailor explanations to individual weaknesses.
\end{quote}

Participants described how encountering incorrect solutions or alternative perspectives can promote deeper engagement and learning. As P11 explained, \textit{``If it shows me an answer that I know it’s wrong, I will stay and think. This will help me stay engaged more'' }. Likewise, P12 highlighted the value of alternative perspectives, suggesting that \textit{``it should provide the user two or three ways of approaching the problem''}. These reflections emphasize the value of contrastive learning, where users engage with correct and incorrect answers, as well as counterarguments that introduce alternative perspectives. This requirement is supported by research showing that engaging with contrasting information and counterarguments can strengthen understanding and reasoning. For instance, studies on argument–counterargument integration show that actively addressing opposing viewpoints improves students’ writing quality and critical thinking~\cite{nussbaum2007promoting}. Building on this evidence,~\citet{wang2025effect} advocate for HOTS-scaffolding (Higher-Order Thinking Skills) to mitigate the lower impact of AI on higher-order thinking skills.

\begin{quote}
    \textbf{DR2: Contrast and counterargument.} The system should combine contrastive questions and counterarguments, helping users strengthen their critical thinking by reflecting on differences and questioning their understanding.
\end{quote}

Participants also expressed a preference for guided, step-by-step learning rather than immediate answers. P3 noted, \textit{``It would be nice to have an option to see the answer, but not from the beginning''}, and P2 emphasized that, \textit{``Human reasoning and critical thinking shouldn’t be diminished''}. Users saw value in a teacher-like system that offers hints, provides feedback, and delivers full solutions only after the learner has attempted the problem. This guided approach was described as more engaging than simply using tools that provide instant answers, as P1 said \textit{``I think it's okay to take a little bit more time than having it be done by someone else''}. This design requirement is informed by research on guided instruction, which shows that learners benefit more from structured guidance than from discovering solutions on their own~\cite{alfieri2011does,kirschner2010minimal}. It also addresses the ``instant answer trap'' identified in usage rate studies~\cite{babu2024chatgpt}.

\begin{quote}
    \textbf{DR3: Guided, step-by-step support.} The system should provide hints and guidance for solving problems, allowing learners to attempt solutions before revealing full answers, to support critical thinking and sustain engagement.
\end{quote}

Finally, participants stressed the importance of dynamic quizzes and a clear, well-organized presentation. Users noted that requiring answers before moving forward would help them fill knowledge gaps, as P5 said, \textit{``It's not about going further into the details, but quizzing me on the current information to make sure I don't have gaps''}. P4 and P5 added, \textit{``If I answer something wrong, it explains and gives further questions''}. P5 also noted the necessity of dynamic questioning, \textit{``Not having the opportunity to answer dynamically the questions adds an unnecessary friction to my work''}. Clear bullet-point summaries were preferred to help retain key information, with P14 adding, \textit{``So that you can absorb the most important things''} and P8 saying, \textit{``If I had a short summary, it would help me remember everything''}. This requirement is supported by prior work on active recall and formative assessment, which demonstrate that quizzes and feedback promote deeper learning and retention~\cite{roediger2006test,butler2008feedback}, and addresses the ``efficiency'' needs highlighted in a survey~\cite{ScholarshipOwl2025GenZAI}, which found that 56\% of Gen Z students use AI for test preparation.

\begin{quote}
    \textbf{DR4: Dynamic quizzing and clear presentation.} This design requirement focuses on the presentation of the system. The system should provide interactive quizzes with feedback and concise bullet-point summaries, making the tool easier to use while promoting active engagement.
\end{quote}
 
\section{Deep3: A Tool for Deliberate, Critical and Analytical Reflection}
\label{sec:App}
\begin{figure*}
    \centering
    \includegraphics[width=1\linewidth]{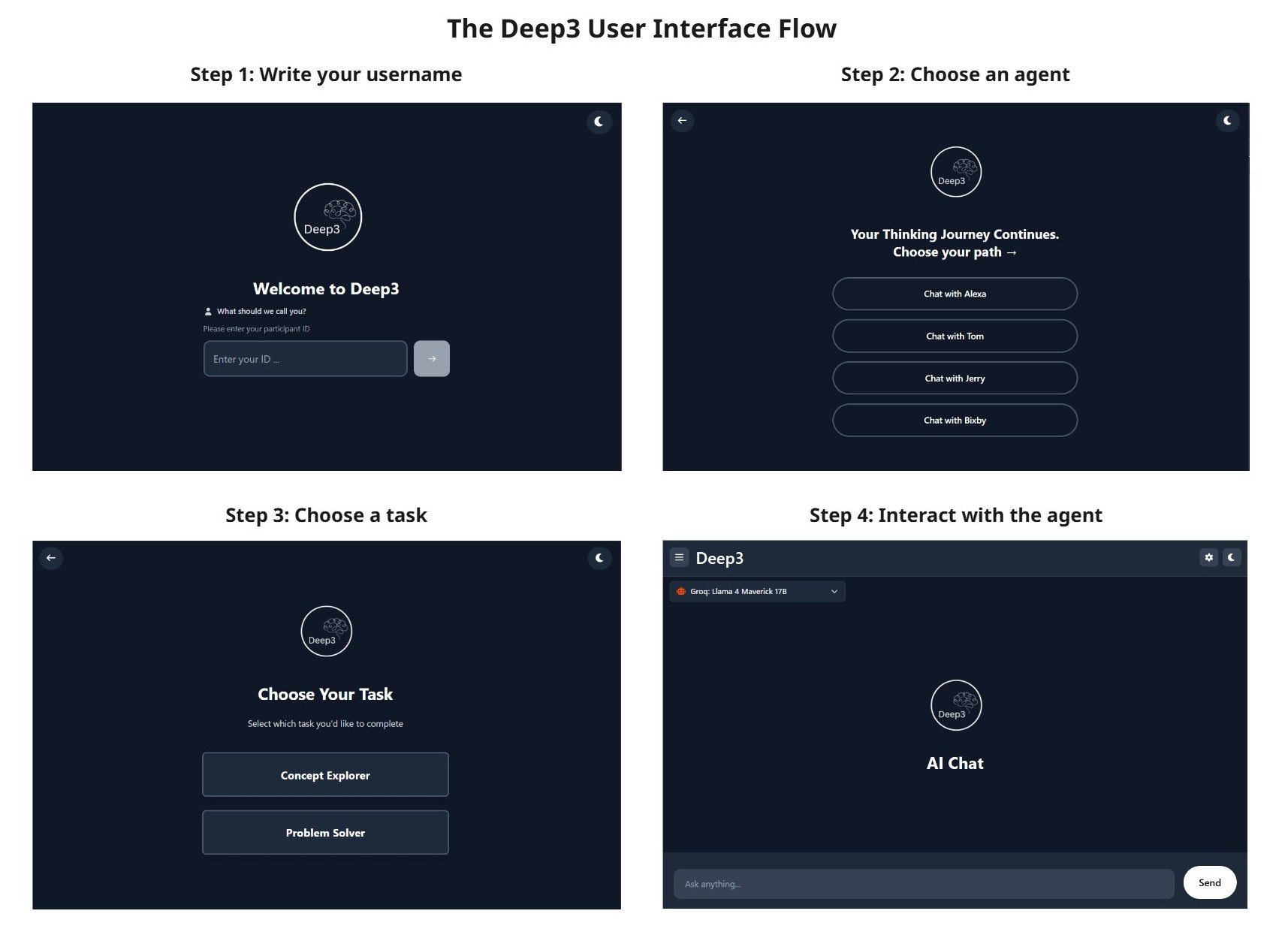}
    \caption{The Deep3 User Interface Flow. The sequence illustrates the four-step onboarding process: (1) user identification via participant ID, (2) selection of a specific AI agent, (3) task type categorization, and (4) the final interactive LLM chat interface.}
    \label{fig:deep3ui}
\end{figure*}

\begin{figure*}
    \centering
    \includegraphics[width=1\linewidth]{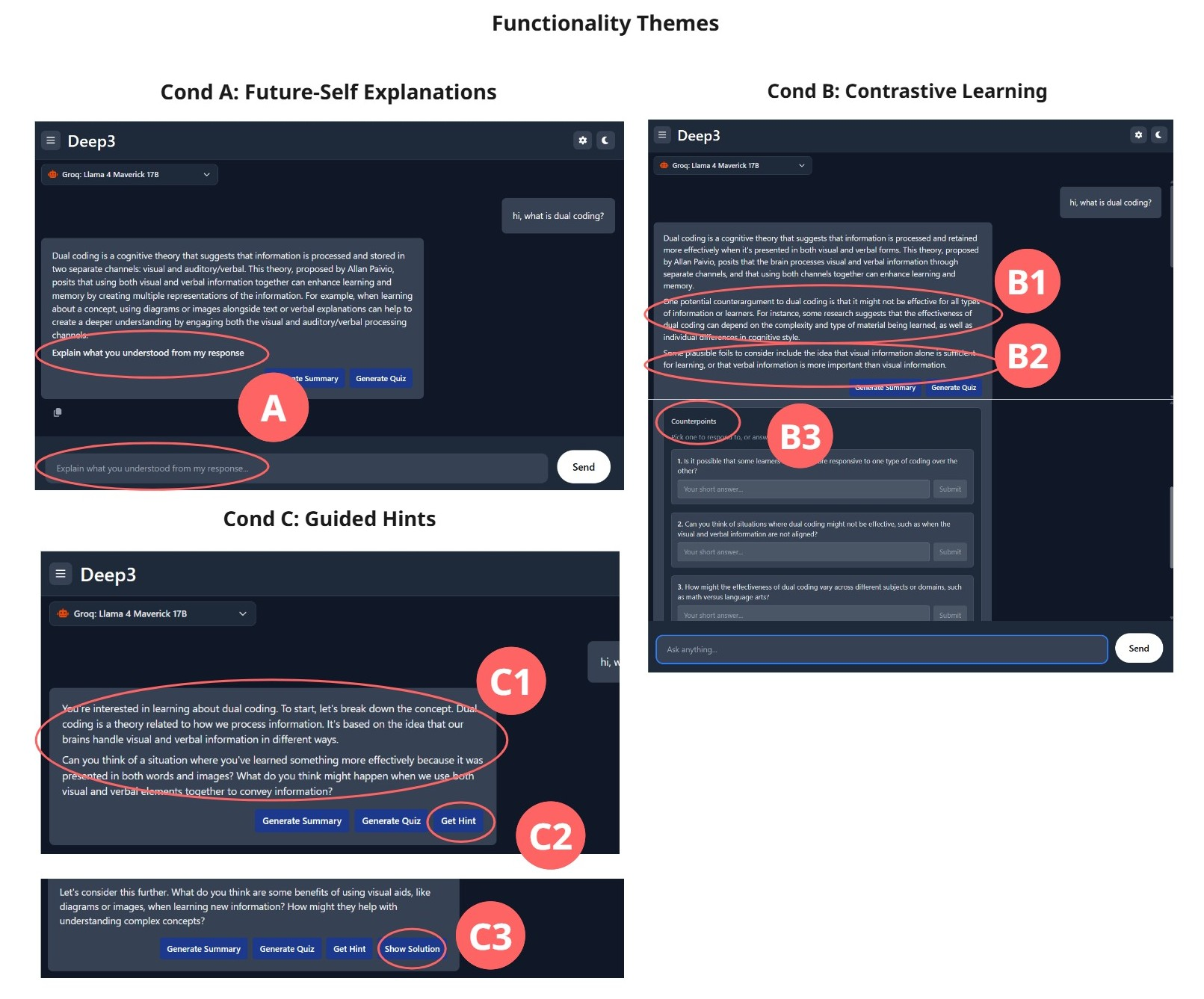}
    \caption{The Functionality Themes. This layout details the specific interactive elements for (A) Cond A: Future-Self Explanations, (B) Cond B: Contrastive Learning, and (C) Cond C: Guided Hints.
    Cond A, focuses on self-explanation through a recurring prompt, ``Explain what you understood from my response'', that ensures the user can accurately describe the topic before proceeding. Cond B, provides a multi-faceted approach involving counterarguments (B1), foils (B2), and specific contrastive questions (B3). Cond C, offers a scaffolded experience where the system breaks down the topic (C1) and provides manual controls for hints (C2) or full solutions after some interaction (C3).}
    \label{fig:conditions}
\end{figure*}

\begin{figure*}
    \centering
    \includegraphics[width=1\linewidth]{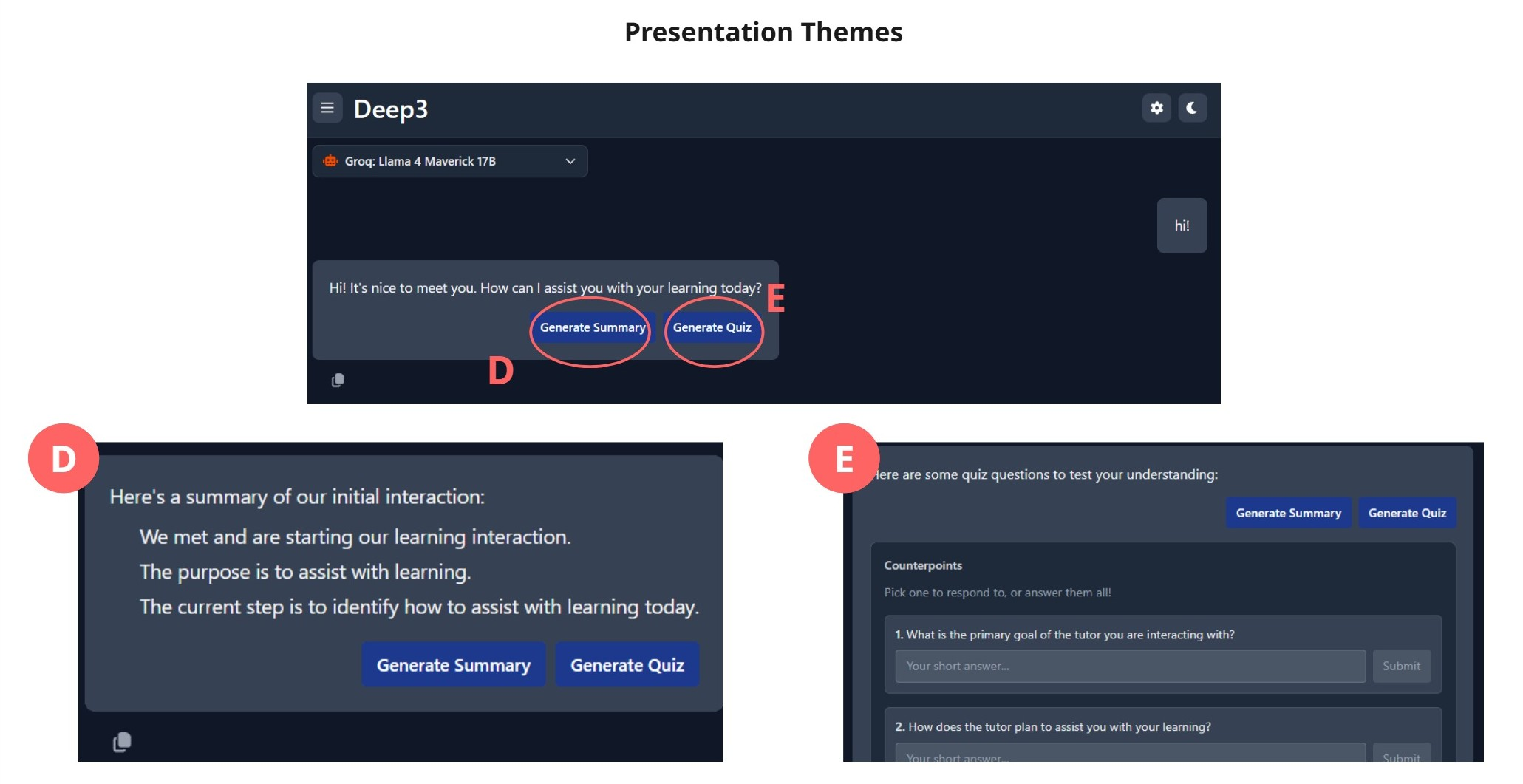}
    \caption{The Presentation Themes. Users have access to tools for generating summaries (D) and dynamic quizzes (E).}
    \label{fig:conditions}
\end{figure*}

Based on the four design requirements identified in our formative study, we developed Deep3 (Figure \ref{fig:deep3ui}). It is a prototype full-stack web application designed to support active reflection and critical thinking during knowledge work. The app offers three different AI agent configurations: Cond A - Future-Self Explanations (\autoref{subsub: conda}), Cond B - Contrastive Learning (\autoref{subsub:condb}) and Cond C - Guided Hints (\autoref{subsub:condc}) addressing, DR1 (\revision{Adaptive Personalization}), DR2 (Contrast and counterargument) and DR3 (Guided, step-by-step support).

\subsection{Functionality Themes: The Three Conditions} \label{sub:conditions}
Deep3 includes three functionality themes: Cond A - Future-Self Explanations, Cond B - Contrastive Learning and Cond C - Guided Hints (Figure \ref{fig:conditions}). To address DR1 (\revision{Adaptive Personalization}), DR2 (Contrast and counterargument) and DR3 (Guided, step-by-step support), the app allows users to choose one of the three functionalities to work with. In this context, ``functionality themes'' refer to the specific interaction styles and behaviors the AI agent uses to help the user.

\subsubsection{Cond A: Future-Self Explanations.}\label{subsub: conda} The first condition is designed to transform the user from a passive reader of AI-generated content into an active participant. In this mode, after the agent provides an answer or explanation, it prompts the user to rephrase and explain that information in their own words. The agent, then, provides feedback to the user and decides if it is necessary to ask for an explanation again.

The theoretical foundation for this condition is rooted in the Self-Explanation Effect. According to \citet{chi1989self}, self-explanation is a critical metacognitive activity that helps learners build a deeper grasp of ``when and why'' certain solutions apply, rather than just memorizing steps.  Furthermore, \citet{mceldoon2013self} suggest that this practice encourages learners to focus on the structural features of a problem rather than superficial details. 

This condition addresses DR1 - \revision{Adaptive Personalization}. By requiring the user to explain concepts in their own words, the system naturally adapts to the user's specific knowledge level (K) and learning style (L) \cite{alshammari2019effective}. If the system detects a misunderstanding in the user’s self-explanation, it immediately adapts its next response to correct those misconceptions.

\subsubsection{Cond B: Contrastive Learning.}\label{subsub:condb} 
The second condition is designed to stop users from simply agreeing with the LLM without thinking it through. In this mode, users cannot just accept an answer and move on, instead they can respond to AI-generated counterarguments or ``foils'' (plausible alternative choice or common misconception the user might have). This interaction gives the option to the user to defend their reasoning or reconstruct their perspective with an opposing view, fostering a more critical and integrative thinking process.

The theoretical foundation for this condition is built on Contrastive Learning. According to \citet{buccinca2025contrastive}, humans naturally think contrastively, often asking ``Why this instead of that?''. We leverage this by predicting a ``foil'' and highlighting the specific differences between the AI’s suggestion and that foil. By making these differences clear and forcing the user to address them, the system helps the user spot their own misunderstandings and refine their thinking.

This condition addresses DR2 - Contrast and counterargument. By making the user to engage with a counterargument, the system creates productive friction that slows down the user's decision-making process. As \citet{nussbaum2007promoting} argue, effective reasoning requires more than just defending a claim, it involves argument-counterargument integration. 

\subsubsection{Cond C: Guided Hints.}\label{subsub:condc} The third condition is designed to help users solve problems without simply handing them the answer. Instead of providing a full solution immediately, the agent acts as a teacher that offers incremental support. The user must actively work through the problem, with the option to unlock hints or the solution only after they have attempted to answer the current step. 

This condition addresses DR3 - Guided, Step-by-Step Support. By breaking down complex information into smaller, manageable pieces, the system ensures the user isn't overwhelmed. According to \citet{alfieri2011does}, ``discovery learning'', where users find answers themselves, is only effective when it is scaffolded. Without clear incremental guidance, users can suffer from ``working memory overload'', which makes it harder to learn. 

Technically, the agent operates using three distinct modes: Default, Hint, and Solution. In the Default mode, the agent ignores direct requests for the final answer and instead nudges the user to think about the next logical step. The ``Hint'' and ``Solution'' modes are only triggered when the user clicks a specific button in the UI. This ensures that the agent remains in a supportive, teaching role.

\subsection{Presentation Themes}
\label{presentation_themes}
In addition to the functionality themes, we implemented Presentation Themes to address DR4 - Dynamic quizzing and clear presentation. While the functionality themes determine what the agent says, the presentation themes control how that information is structured and delivered to the user. This layer is integrated throughout the entire interface, allowing users to transform responses from any of the three previous agents into more effective learning formats, such as summaries or interactive quizzes.

The theoretical backbone of these presentation themes is Test-Enhanced Learning. According to \citet{roediger2006test}, the act of retrieving information from memory is a more powerful learning event than simply re-studying or re-reading material. To apply this, the system can enter Quiz Mode, which triggers the generation of 3-5 short-answer questions. Short-answer questions were chosen based on the literature's recommendation that the cognitive advantage of retrieval is significantly stronger for recall tasks than for multiple-choice tasks.

For moments when the user needs to organize their thoughts, the system offers Summary Mode. Based on the literature's recommendation to present material ``clearly and concisely'', this mode reformats complex AI responses into structured bullet points. 

\subsection{User Interface} The Deep3 interface was designed to follow a familiar chat-based paradigm, similar to standard LLM interfaces. This choice was intentional to minimize the ``learning curve'', allowing users to focus entirely on their tasks rather than navigating a complex new tool. 

While the interface looks like a standard chatbot, it includes several custom components to support the Functionality and Presentation Themes. For example, the UI includes interactive elements like mode-switching buttons, to manually trigger ``Quiz Mode'' or ``Summary Mode'' (Figure \ref{fig:conditions}). 

\subsection{Platform and Implementation} Deep3 is implemented as a full-stack web application. The frontend is built using React 19 and Vite, providing a responsive user interface, with Tailwind CSS for styling. We integrated React Markdown for rich-text rendering and FontAwesome for consistent icons throughout the app.

The backend is built with FastAPI, a Python web framework that handles the logic and connects to our database. We used MongoDB to store user data that was gathered in our summative study. To manage the different AI models, we integrated any-llm \cite{mozillaAnyLLM}, an open-source library that provides a standard way to connect to various LLM providers. For this prototype, we used Groq development key to get responses from the llama-4-maverick-17b-128e-instruct model. Our choice was informed by the LMSYS Chatbot Arena leaderboard \cite{chiang2024chatbot}, where this model maintained a top position among available models for text-based reasoning.

For deployment and environment consistency, the backend is containerized using Docker. The application is hosted in a distributed environment, with the frontend deployed on \href{https://vercel.com/}{vercel.com} and the backend managed through \href{https://render.com/}{render.com}.

\subsection{Prompt Engineering} To ensure that the agents provide consistent and helpful feedback, we followed established best practices for prompt engineering techniques. Our approach was guided by several techniques, specifically focusing on providing clear instructions, explicit constraints, and relevant context to improve the model's reasoning performance~\cite{ekin2023prompt,schulhoff2024prompt}.

Every prompt used in Deep3 follows a standardized three-part structure to maintain consistency: \textit{Role, Context, and Task}. First, we assign the agent a specific Role, to establish the tone and perspective of the response. Next, we provide the theoretical background for the specific reflection technique being used. This information is based on established educational literature and ensures the agent stays grounded in proven pedagogical methods. Finally, the Task defines the direct instructions the model must follow. Every task begins by requiring the agent to strictly take into account the theoretical context described in the prompt for every response. From there, the task defines specific functional behaviors for every condition. \revision{The complete set of these system prompts is provided in Appendix \ref{app:prompt}}.

\section{User Evaluation}
\label{sec:user-evaluation}

\subsection{Methodology}
The evaluation study was conducted \revision{between December 2025 and January 2026}  as a between-subjects study~\cite{charness2012experimental}, consisting of two phases: a main study session, and a follow-up interview phase for a subset 14\% of the sample. The first phase, was designed to allow participants to engage with the Deep3 application, which included two tasks, each implemented across four different agent conditions. Participants were randomly assigned to one condition to avoid cross-contamination of results. The sessions, including task completion and questionnaire responses, lasted between $25-40$ minutes, depending on the participant. To ensure ecological validity while maintaining a controlled environment, all sessions were conducted remotely via Microsoft Teams. This allowed participants to use their own laptops in a familiar setting while ensuring minimal external distractions. In the second phase, 14\% of the participants took part in a semi-structured interview lasting approximately $10-20$ minutes. We used a mixed-methods approach~\cite{creswell1999mixed}, combining quantitative data from questionnaires with qualitative insights from both interviews and log files recorded during the Deep3 interaction phase. The study was approved by \revision{the Cyprus National Bioethics Committee.}

\subsection{Pilot Studies}

To ensure the validity and effectiveness of the experimental design, six pilot studies were conducted~\cite{van2002importance}. These sessions served to refine the protocol and adjust the task difficulty. During the first three pilot sessions, the second task was initially designed as a scheduling problem. However, participant feedback revealed that the task felt more like an organizational exercise than a genuine problem-solving challenge, and they noted a preference for completing the task manually rather than engaging with the AI.

Consequently, for the remaining three pilot sessions, the second task was redesigned into a more complex, problem-oriented task that required higher-level reasoning. This adjustment ensured that the task was better suited to evaluate the utility of the AI intervention. Furthermore, these six preliminary studies were used to empirically determine the appropriate duration for each task, ensuring that the time limits were sufficient for completion without being excessive.

\begin{table*}[t!]
    \centering
    \caption{Description of behavioral metrics and subjective outcome variables.}
    \label{tab:logs}
    \scalebox{0.85}{
    \begin{tabular}{lllll}
    \toprule
    \textbf{Variable} & \textbf{Description}\\ 
    \midrule
    \textit{General Behavioral Metrics} &\\
    Total Prompts & Total number of prompts generated by the user to complete the task\\
    Task Completion Time & Total time (minutes) taken to finish the task\\
    Average Response Time & Mean duration (seconds) for a user to respond to system prompts\\
    Summaries Count & The count of summaries accessed\\
    Quiz Count & The number of quizzes generated\\
    \midrule
    \textit{Learning \& Subjective Outcomes} &\\
    Actual Understanding & Mean scores for the user's actual understanding\\
    Subjective Measures & Mean item scores for each of the four post-experiment questionnaires\\
    \midrule
    \textit{Condition-Specific Measures} &\\
    Self-explanations & Number of self-explanations made by the user (Cond A)\\
    Counterargument engagement & Number of counterpoint quizzes taken and the ratio of questions answered (Cond B)\\
    Hints and Solutions & Total number of hints requested and full solutions accessed (Cond C)\\
    
    \bottomrule
    \end{tabular}
    }
\end{table*}

\subsection{Experimental Tasks}

To ensure ecological validity, the tasks were designed to align with documented LLM usage patterns among university students. This research shows that students often use LLMs to clarify academic concepts and assist in problem-solving or homework completion \cite{paustian2024students, ravvselj2025higher,freeman2025student}. Consequently, two distinct tasks were developed.

\textit{Task 1: Concept Explanation.} This task focused on academic clarification. Participants were given a university-related concept, and they were asked to engage with the chatbot until they felt they understood that concept. Based on our pilot studies, participants were given 6-8 minutes to finish this task, with the specific duration adjusted according to each participant’s English proficiency.

\textit{Task 2: Problem Solving.} This task required participants to solve a logic-based puzzle using the chatbot. This problem involved a balance scale and a set of coins, and an extra constraint to increase the complexity of the problem.  Participants were given 10-15 minutes for this task, to accommodate varying levels of English proficiency.

Both tasks were presented in a randomized order to mitigate potential fatigue or learning effects. While maximum durations were set based on pilot data, participants were permitted to conclude a task early if they believed they had successfully reached a solution or achieved a sufficient understanding of the concept. \revision{Both tasks can be found in Appendix~\ref{app:tasks}.}

\subsection{Procedure and Data Collection} 

Participants signed up for the study by booking a time slot based on their availability. Depending on whether they volunteered for just the main interaction or both the interaction and the follow-up interview, they were allocated either a 40-minute or a 60-minute session. Upon joining the Microsoft Teams meeting, they first completed a consent form.  They then completed a demographics survey, which included questions about their previous experience with LLMs in an educational context. A researcher then introduced them to the Deep3 tool and provided a short walkthrough of the interface, creating a low-pressure environment for asking questions and getting familiar with the interface. 

Participants then proceeded with the experimental tasks, interacting with one of the four agent conditions. Upon completing the task, they filled two questionnaires: (1) understanding assessment questionnaire, which consisted of a perceived understanding question, using a 5-point Likert scale, and an actual understanding open-ended question which was later graded by a researcher using a predetermined marking scheme \cite{nimmo2024user}; and (2) the NASA-TLX questionnaire \cite{hart2006nasa}. 

The same structure was followed for Task 2. Throughout the entire session, participants shared their screens via Microsoft Teams. This allowed the researcher to monitor the interaction and provide technical support if any issues arose with the interface. After completing both tasks, participants filled out two final questionnaires, to evaluate their overall impression of the system: (1) the Trust Questionnaire (Appendix D from \cite{hoffman2018metrics}); and (2) the short version of the User Experience Questionnaire (UEQ-S) \cite{laugwitz2008construction}.

The subset of participants who agreed to the follow-up phase then took part in a 10–20 minute semi-structured interview. These interviews focused on their overall impressions of the Deep3 application and their specific thoughts on the agent condition they used. Additionally, participants were asked for suggestions on how to improve the tool and whether they believed it would be a helpful resource for guidance in a school or university setting. \revision{The interview protocol can be found in Appendix~\ref{app:interview}.}

\subsection{Participants}

We recruited 85 participants (P1-P85) through personal networks and advertisements through the authors' institutions. \revision{Participation was entirely voluntary, and all individuals provided informed consent prior to the study.} The sample included individuals from various academic and professional backgrounds, with the most represented fields being Computer Science, Mathematics, Engineering (Civil, Computer, and Chemical), Philology, and Economics. All participants resided in Europe, \revision{and studied in universities in the United Kingdom, the Netherlands, Italy, Bulgaria, Greece, and Cyprus.} Their ages were from 19 to 27 years ($M = 22.14, SD = 1.66$), 41 identified as women, 43 as men, and 1 as other. In terms of education, the majority were currently enrolled in academic programs, including 57 undergraduate, 13 master’s, and 3 doctoral students, while 12 participants had already completed their degree (6 Bachelor's and 6 Master's). All participants were proficient in English, with over $90\%$ reporting a B2 level (Upper Intermediate) or higher. Regarding frequency of interaction with LLMs, participants reported high usage, with a mean general usage score of $3.67$ ($SD = 1.17$) and an educational usage score of $3.86$ ($SD = 1.15$) on a 5-point Likert scale. The most common applications of LLMs in their education included learning and understanding concepts ($90.6\%$), answering homework ($56.5\%$), and summarizing text ($55.3\%$).

\begin{figure*}[t]
\centering\subfloat[Average Response Time (sec)]{\includegraphics[width=3.7cm]{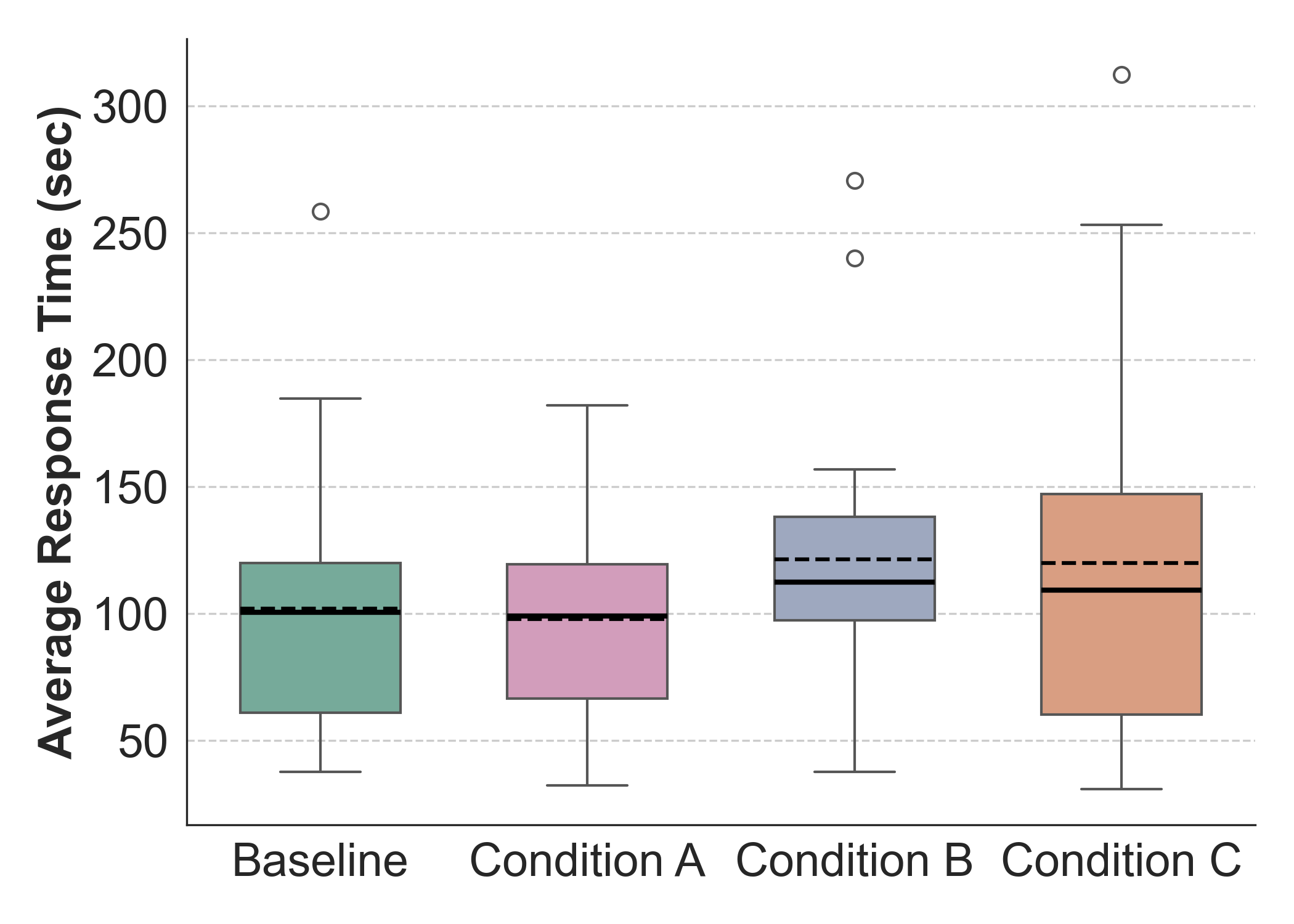}}\hfil
\subfloat[Task Completion Time (min)]{\includegraphics[width=3.7cm]{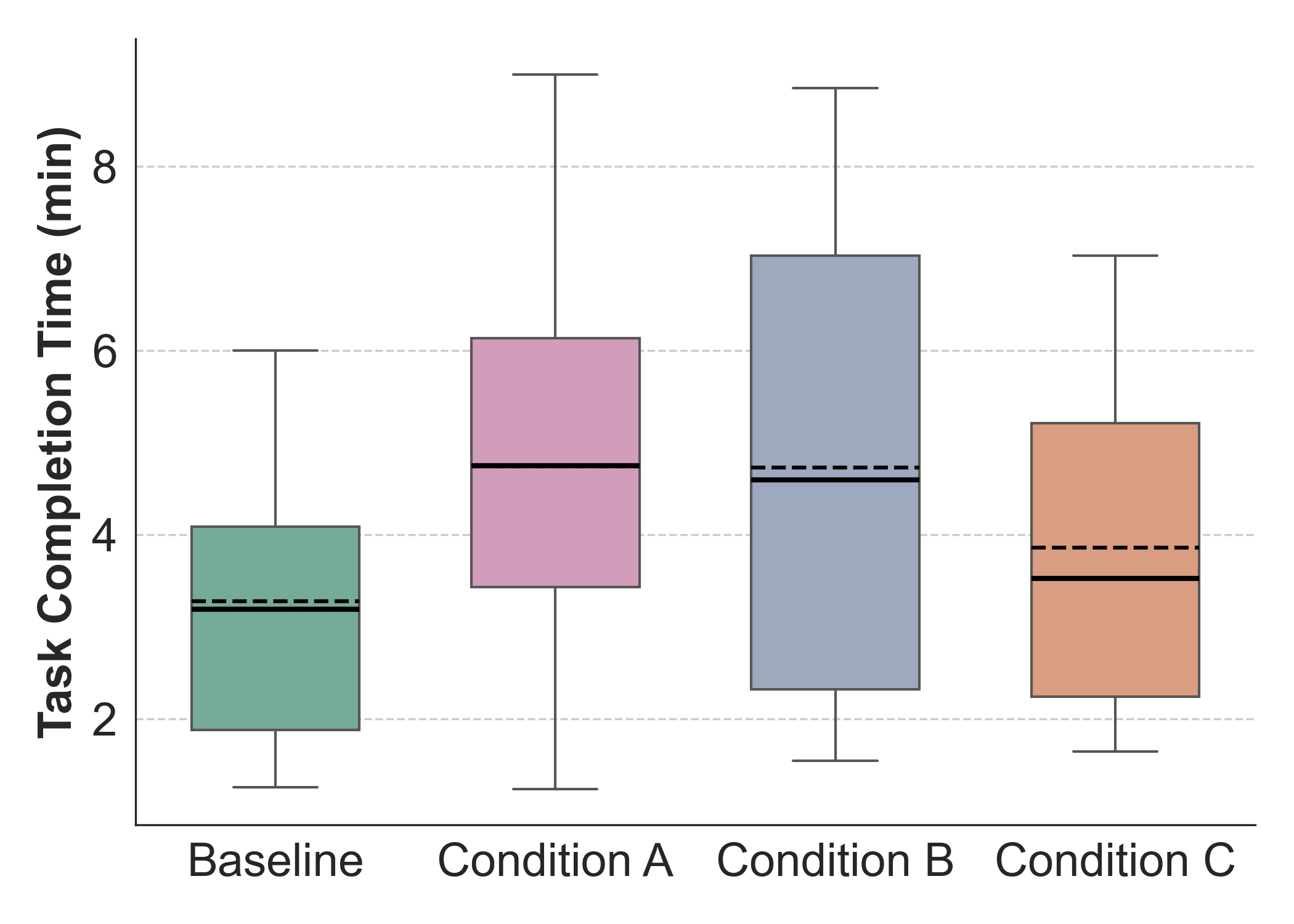}}\hfil \subfloat[Total Prompts]{\includegraphics[width=3.7cm]{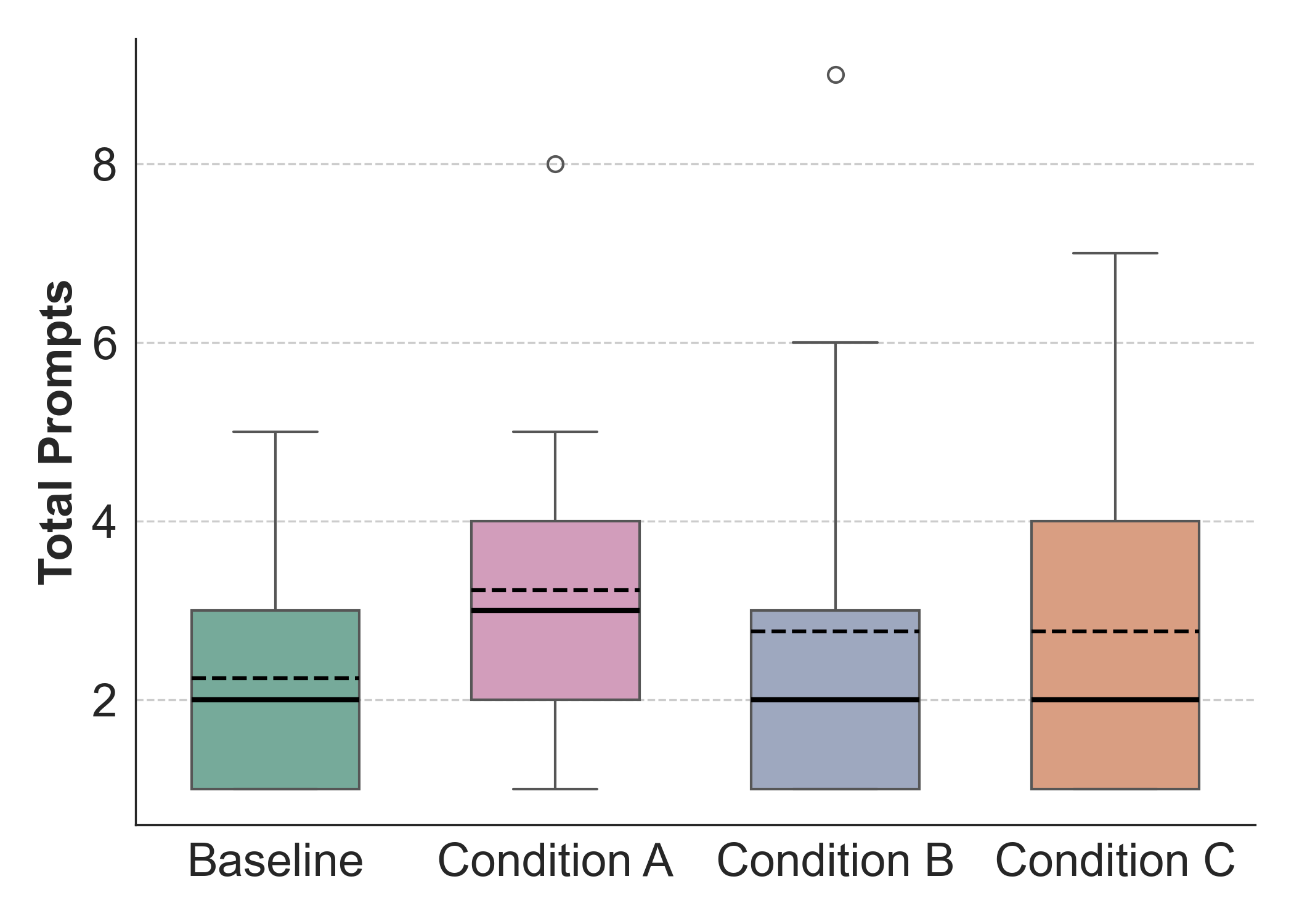}} 
\hfil \subfloat[Task Performance]{\includegraphics[width=3.7cm]{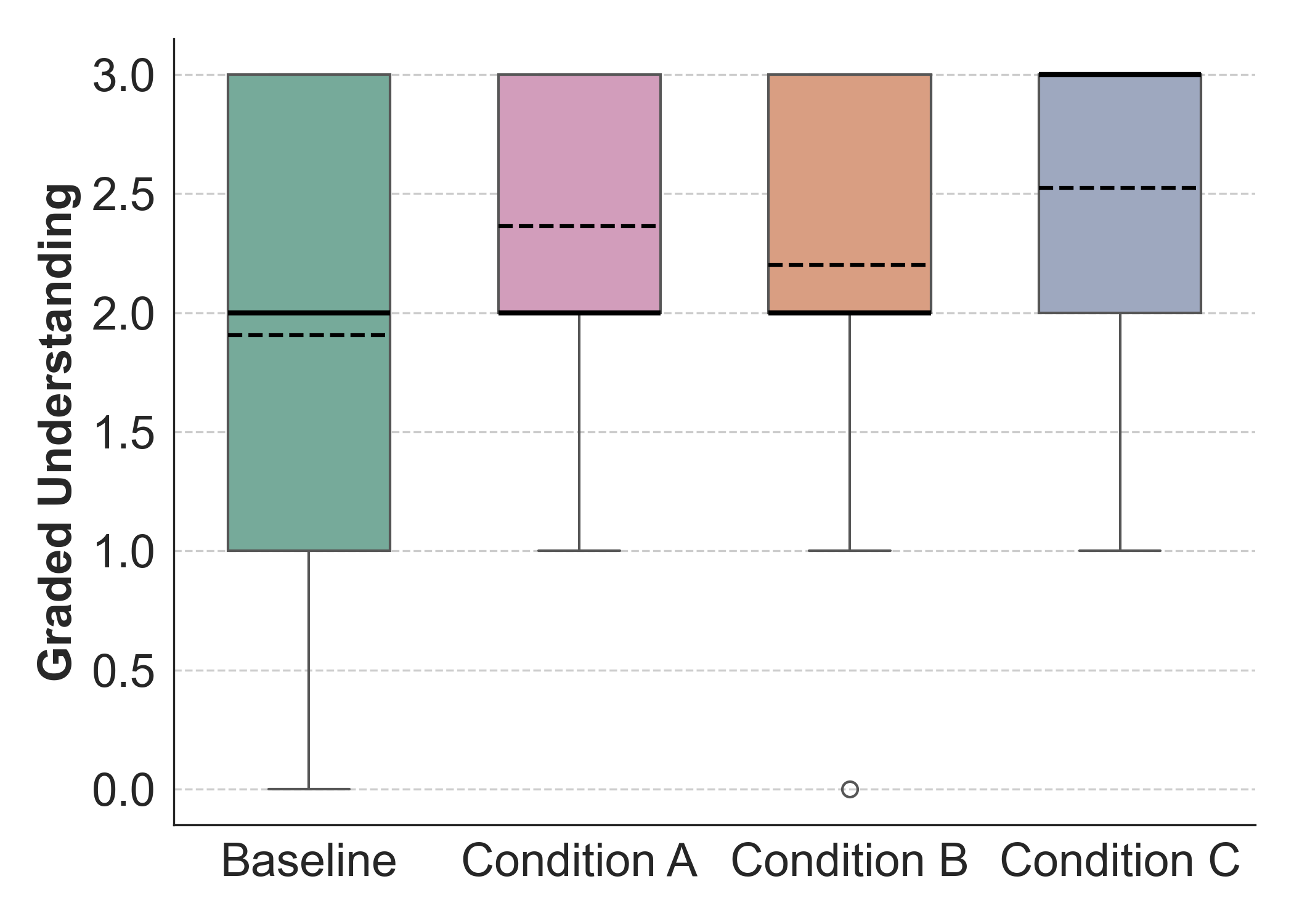}} 
\caption{Task 1: Concept Explanation. Performance across four interaction modes (Baseline, Cond A: Future-Self Explanations, Cond B: Contrastive Learning, Cond C: Guided Hints). Box plots show per-condition distributions across all participants; solid lines denote medians, dashed lines denote means. Note: Task 1 task performance scores range from 0-3.}\label{task1}
\end{figure*}

\begin{figure*}
\centering\subfloat[Average Response Time (sec)]{\includegraphics[width=3.7cm]{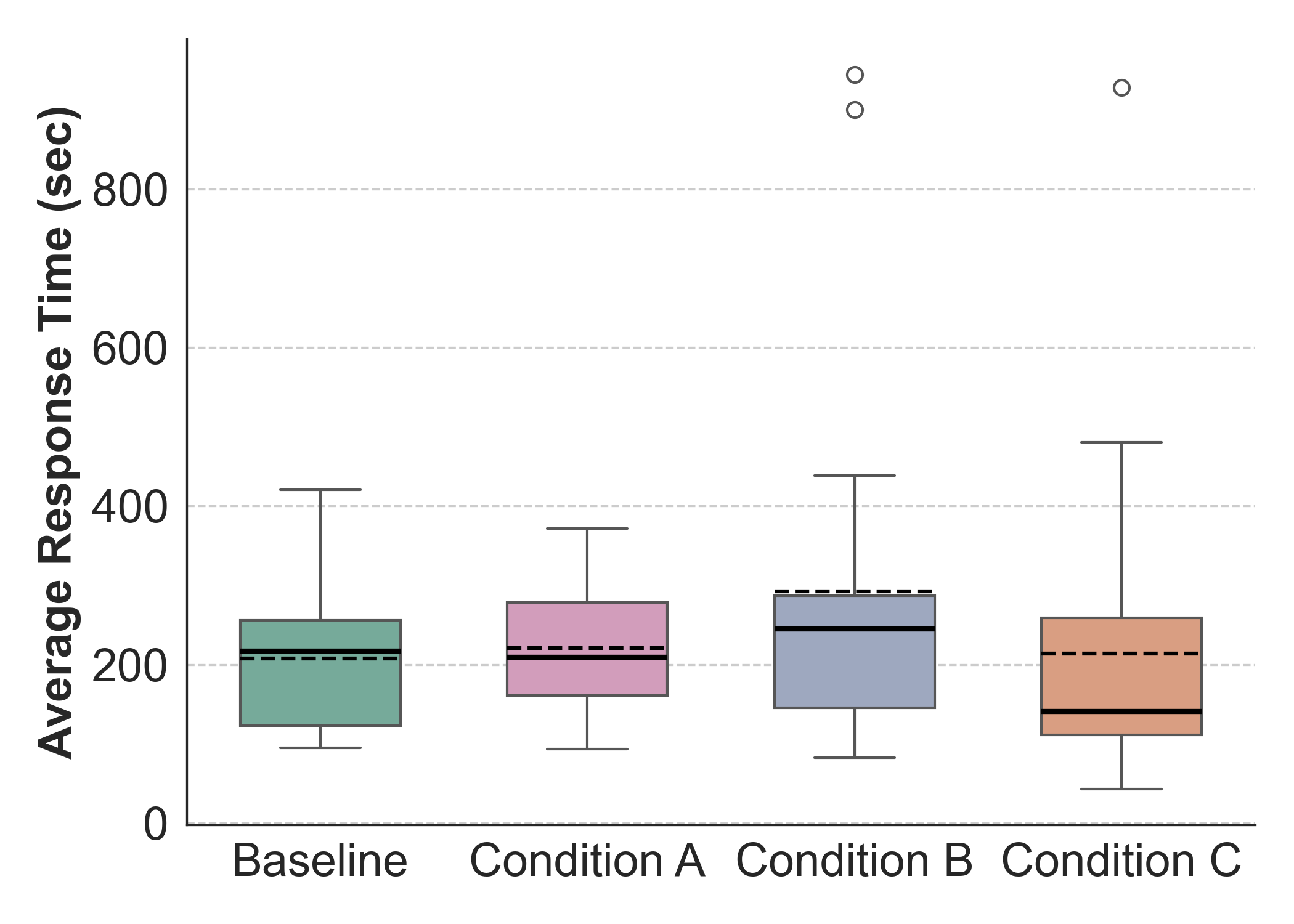}}\hfil \subfloat[Task Completion Time (min)]{\includegraphics[width=3.7cm]{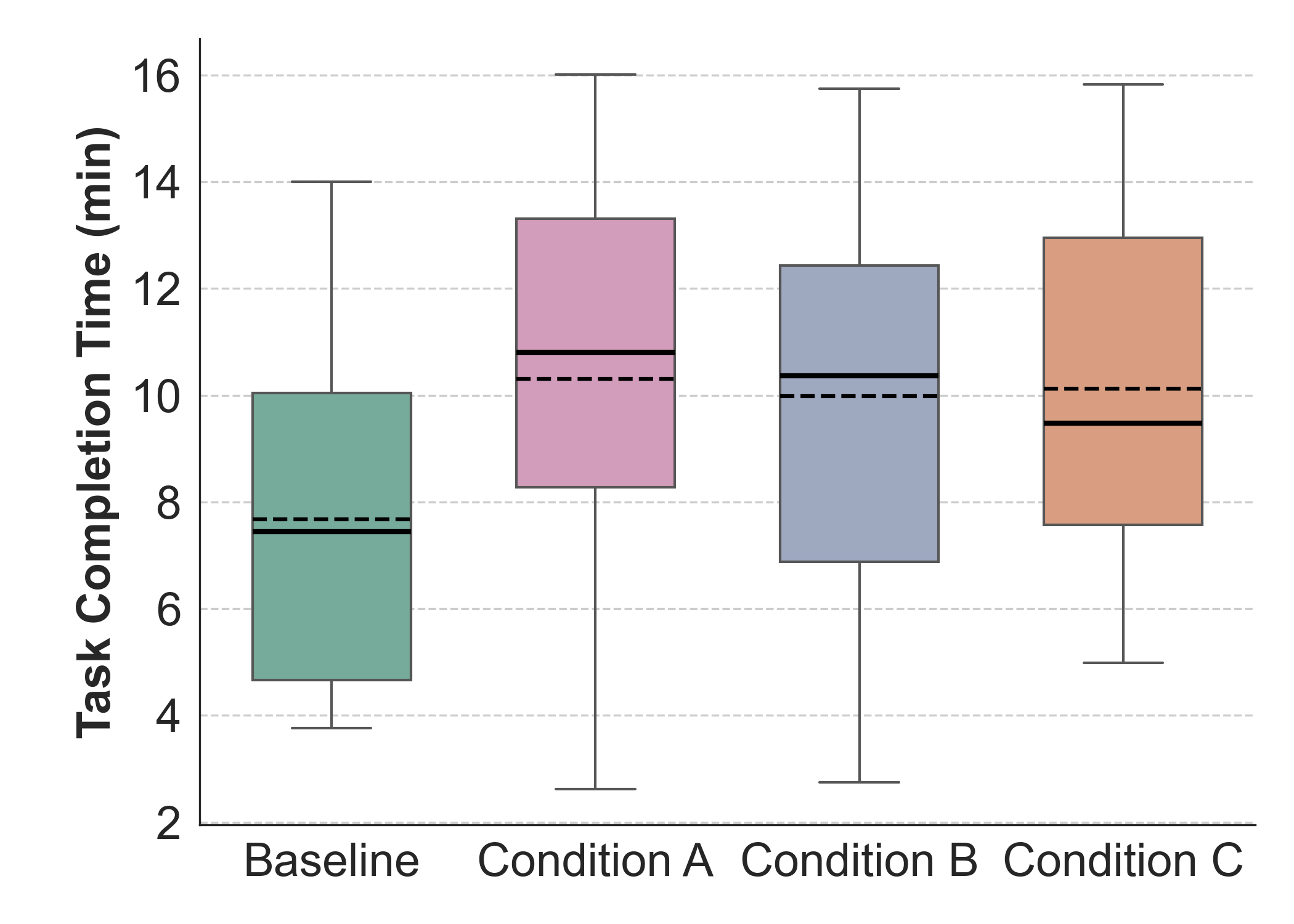}}\hfil
\subfloat[Total Prompts]{\includegraphics[width=3.7cm]{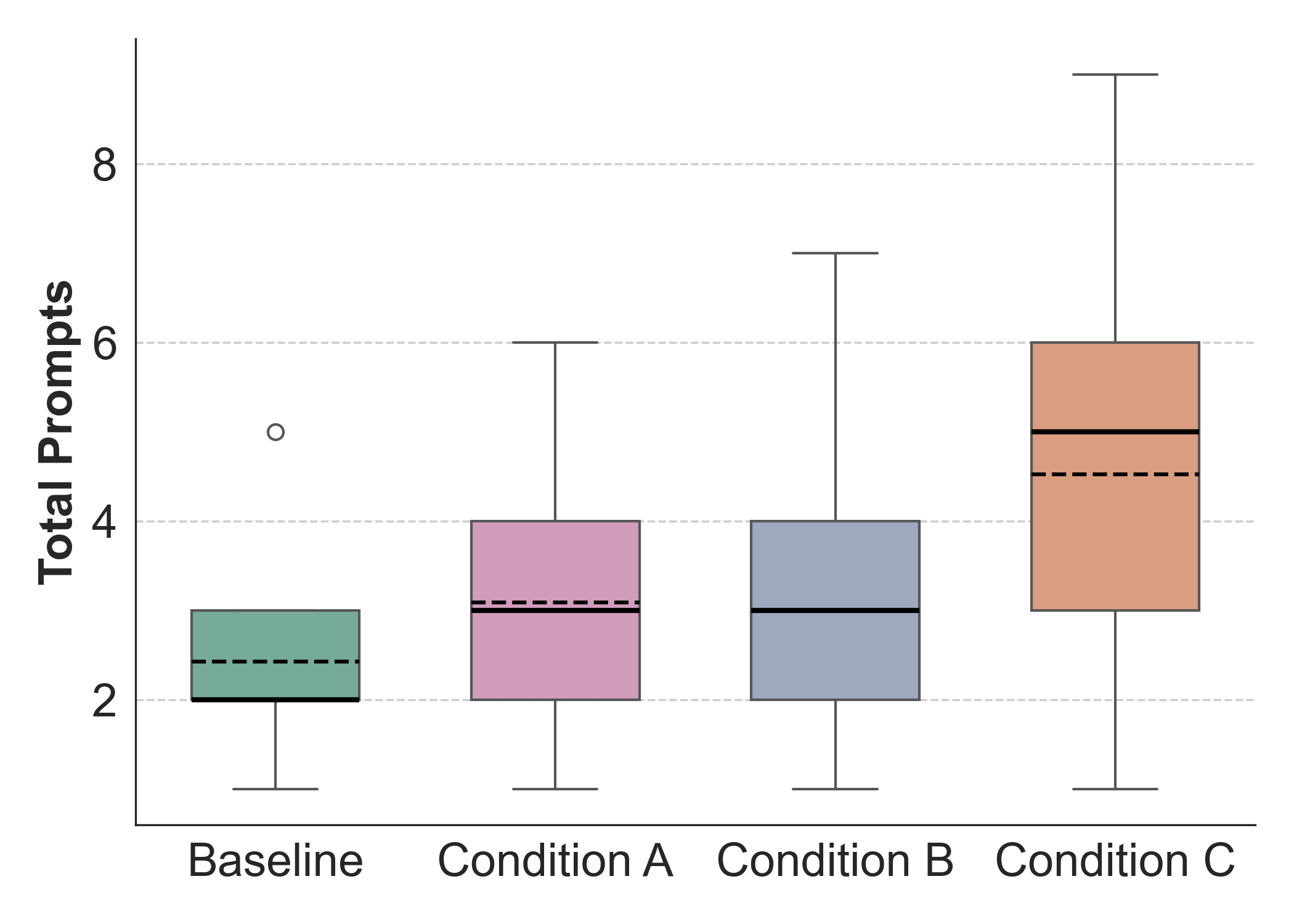}}
\hfil
\subfloat[Task Performance]{\includegraphics[width=3.7cm]{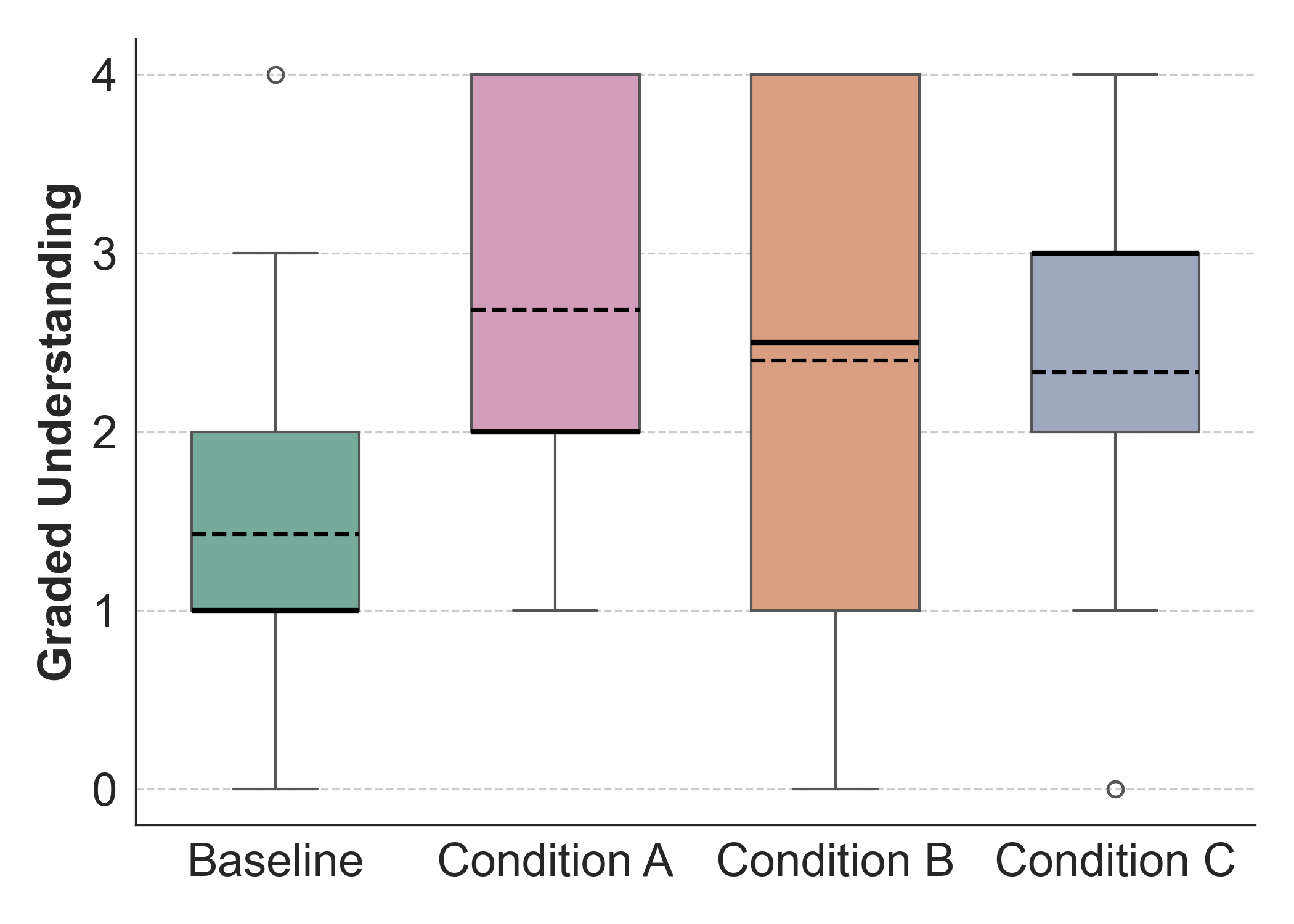}}
\caption{Task 2: Problem Solving. Performance across four interaction modes (Baseline, Cond A: Future-Self Explanations, Cond B: Contrastive Learning, Cond C: Guided Hints). Box plots show per-condition distributions across all participants; solid lines denote medians, dashed lines denote means. Note: Task 2 task performance scores range from 0-4.}\label{task2}
\end{figure*}

\subsection{Data Analysis}

\subsubsection{Quantitative Analysis.} Since the study evaluates user engagement and learning efficiency within the application, performance and user experience were measured across several objective and subjective metrics. We categorized these metrics into general behavioral logs and condition-specific measures adapted to the unique interventions of each study group. Table \ref{tab:logs} provides a comprehensive definition of these variables.

To explore the relationship between user behavior and learning outcomes, we performed a Pearson correlation analysis. Specifically, we examined pairwise correlations between behavioral measures (e.g., time on task) and subjective outcomes (e.g., questionnaire scores), as well as internal correlations among the behavioral measures themselves to identify patterns in user interaction.

For each metric, we assessed normality using Shapiro–Wilk tests across all experimental conditions. When normality assumptions were satisfied, we conducted One-Way ANOVA to evaluate global differences across conditions; otherwise, we applied the non-parametric Kruskal–Wallis H-test. To compare specific conditions, we performed planned pairwise comparisons (Independent T-Tests). For behavioral metrics, we compared each experimental condition against the Baseline condition. For feature-specific metrics (summaries used and quizzes generated), we conducted pairwise comparisons exclusively among the experimental conditions. All pairwise comparisons were Bonferroni-corrected to account for multiple testing. We report p-values for significance and Cohen’s d \cite{cohen2013statistical} for effect sizes, interpreted as small ($>0.2$), medium ($>0.5$), and large ($>0.8$). Additionally, we report eta-squared ($\eta^2$) for ANOVA tests to quantify the proportion of variance explained by condition differences.

\subsection{Qualitative Analysis} All interviews were audio-recorded and then transcribed with \href{https://www.rev.com}{rev.com}. Transcripts were then reviewed to endure there were no inconsistencies. The cleared interviews were qualitatively analyzed using thematic analysis \cite{braun2006using}. We employed an iterative open-ended coding process, identifying data patterns related to our research questions. \revision{First, the lead author independently coded a representative subset of the transcripts. The research team then met to review these initial codes and discuss the developing coding scheme. During this session, the team resolved several coding disagreements, and once a final coding scheme was agreed upon,  the lead author coded the remaining transcripts using \href{https://www.miro.com}{miro.com}.}
Code titles included: \textit{impression of the tool, cognitive depth, perceived pedagogical utility, perceived knowledge retention and active recall interventions}. Through this iterative process, we identified three
themes in relation to our research questions.

\section{Results}
\label{sec:results}

Next, we present the quantitative and qualitative results from our user study. Full statistical details for all analyses are reported in Table~\ref{tab:anova} for performance metrics and Table~\ref{tab:anova2} for subjective measures.

\begin{figure*}
\centering\subfloat[Cond A: Self-Explanations Count]{\includegraphics[width=4.1cm]{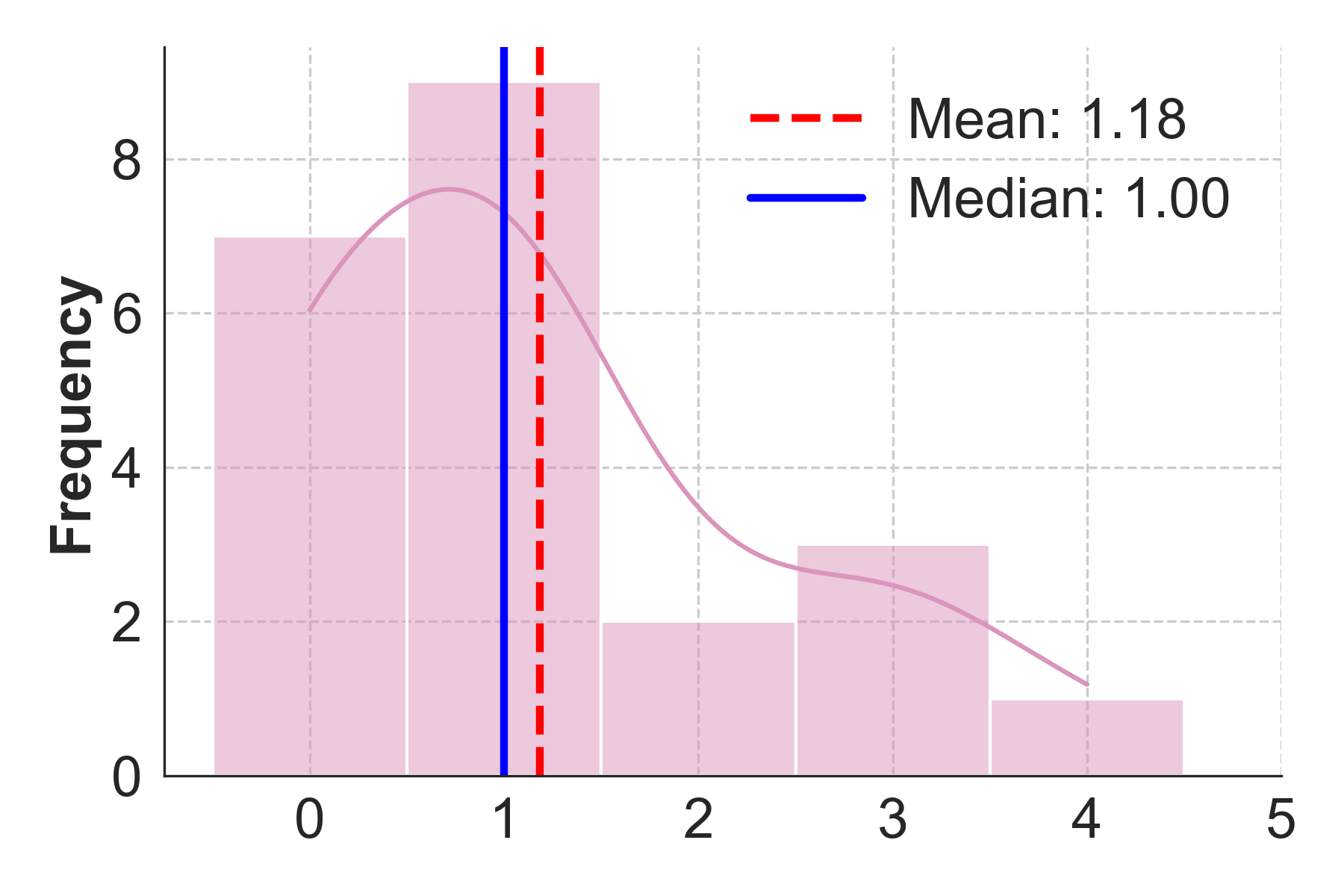}} \subfloat[Cond B: Contrastive Questions Answered]{\includegraphics[width=4.1cm]{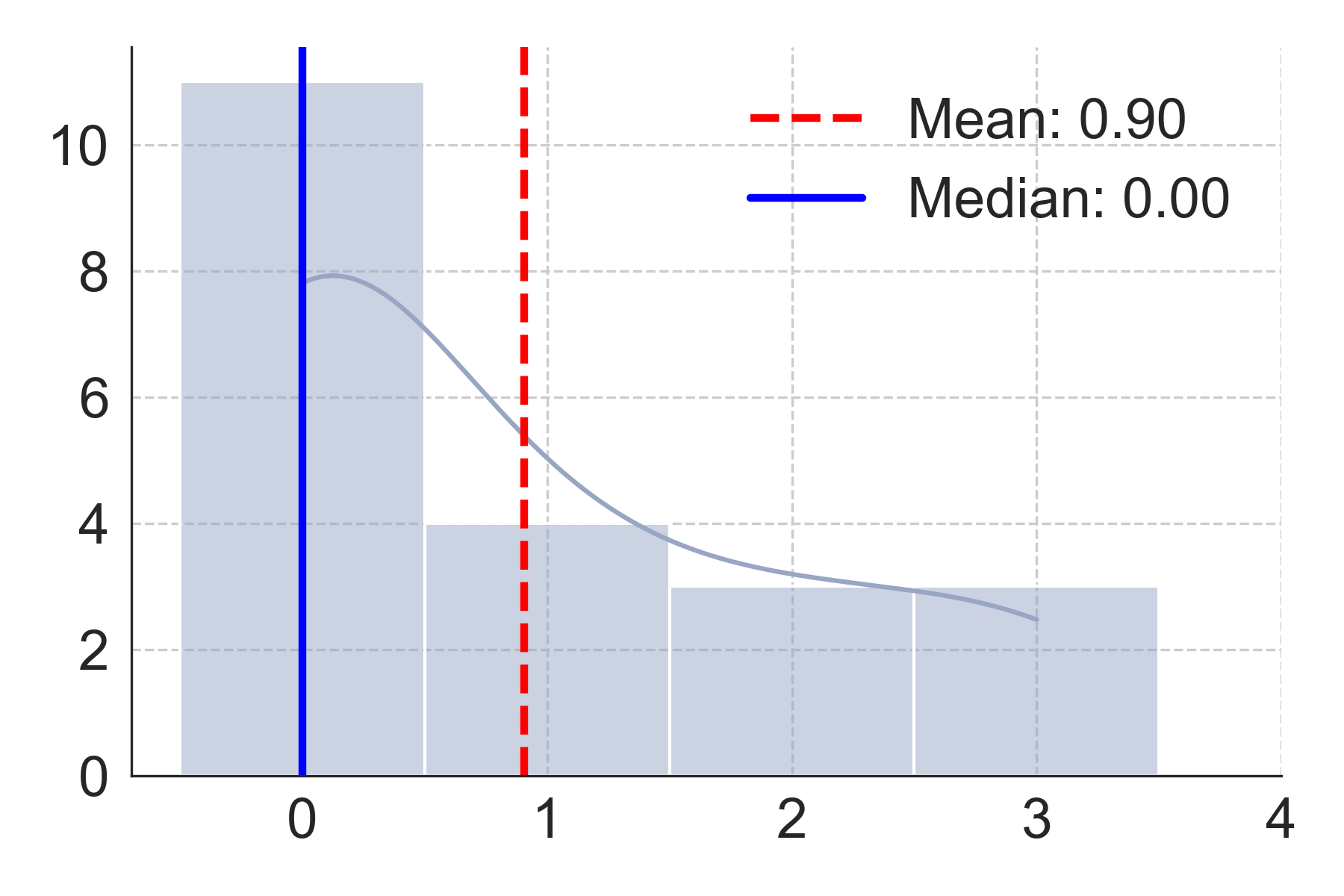}}\subfloat[Cond C: Hints Used]{\includegraphics[width=4.1cm]{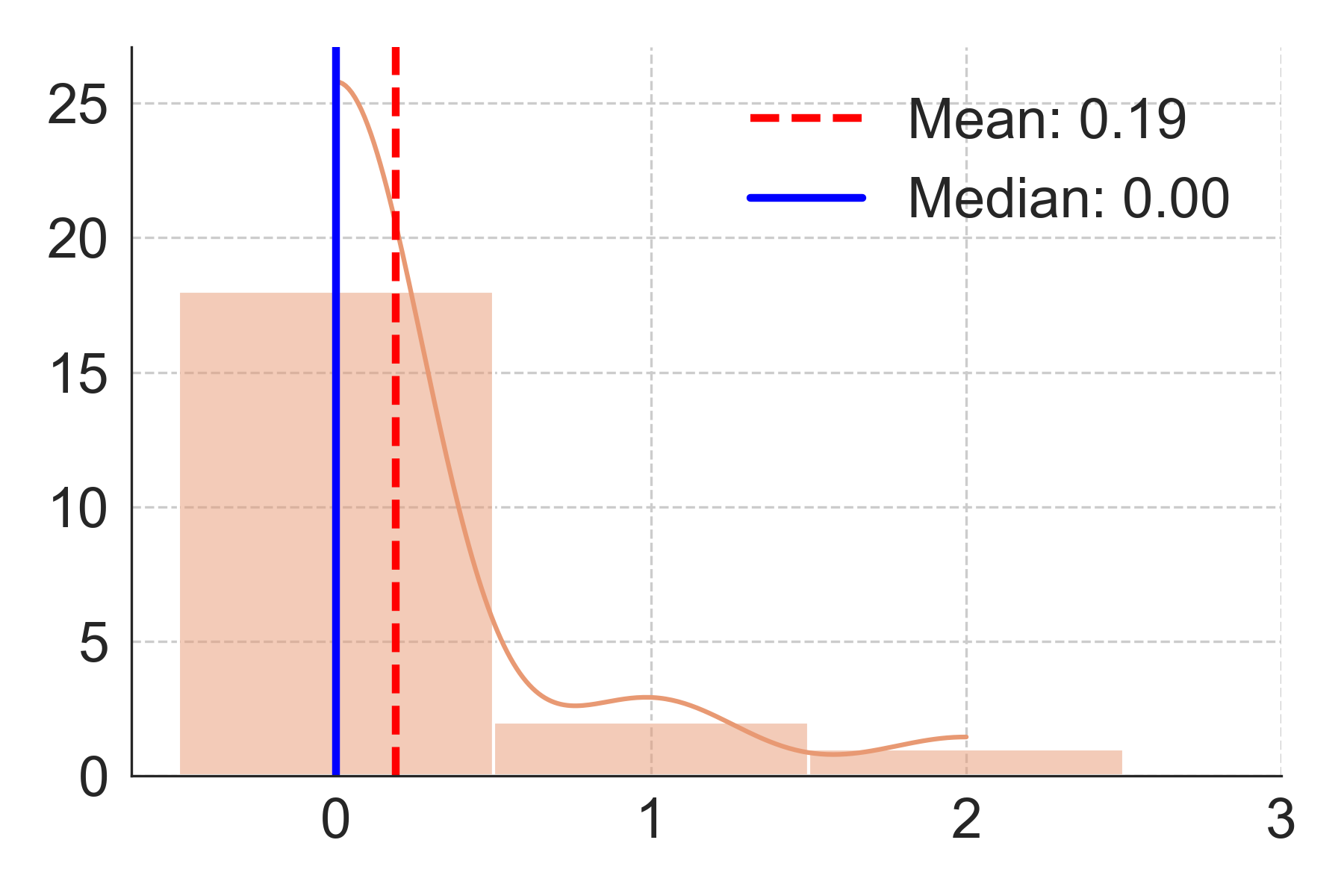}} \subfloat[Cond C: Solutions Used]{\includegraphics[width=4.1cm]{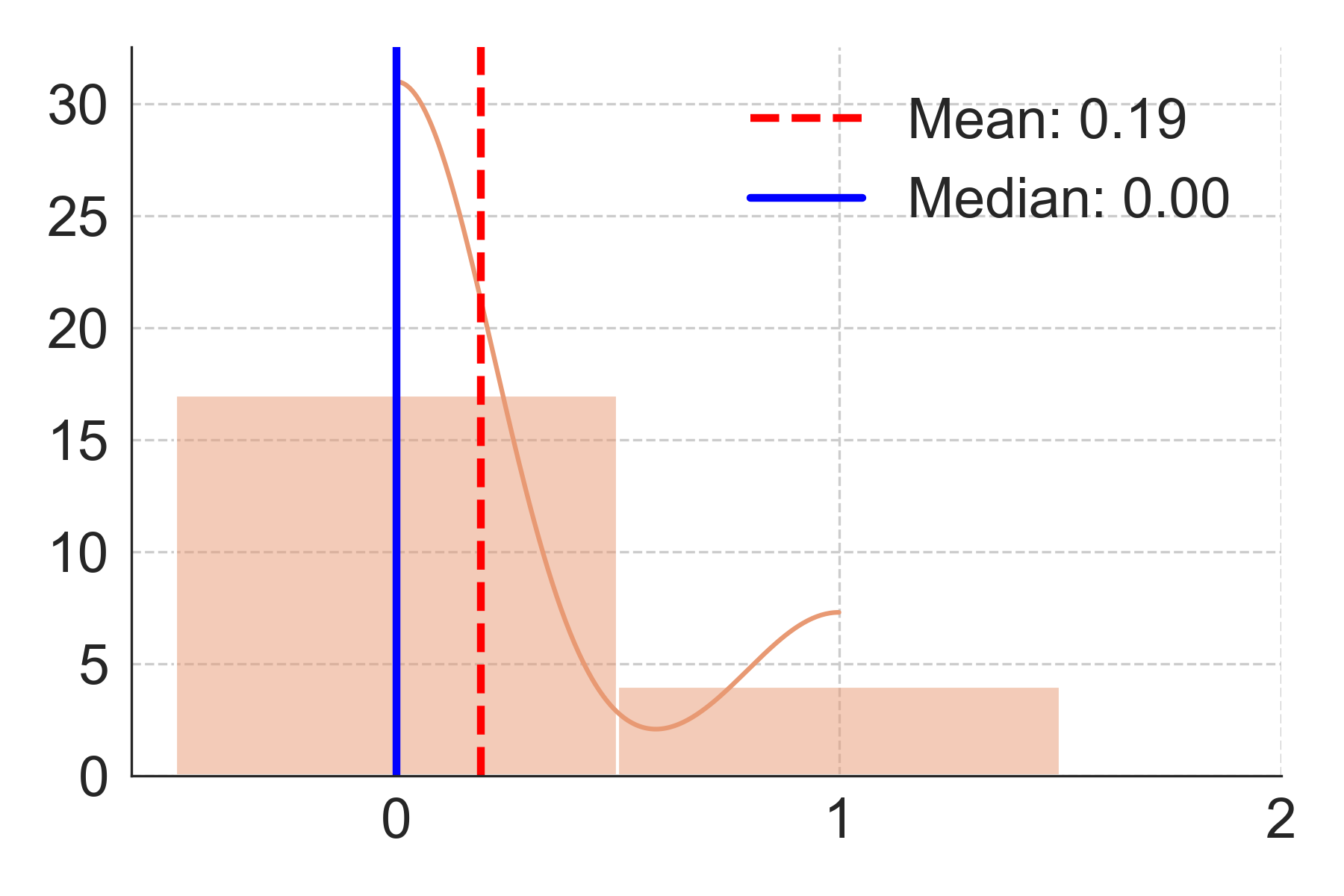}}
\caption{Task 1: Concept Explanation. Condition specific feature usage across three interaction modes (Cond A: Future-Self Explanations, Cond B: Contrastive Learning, Cond C: Guided Hints). Histograms illustrate per-condition distributions across all participants; solid lines denote medians, dashed lines denote means, and the curve represents the kernel density estimation.}\label{conditionspecific1}
\end{figure*}

\begin{figure*}
\centering\subfloat[Cond A: Self-Explanations Count]{\includegraphics[width=4.1cm]{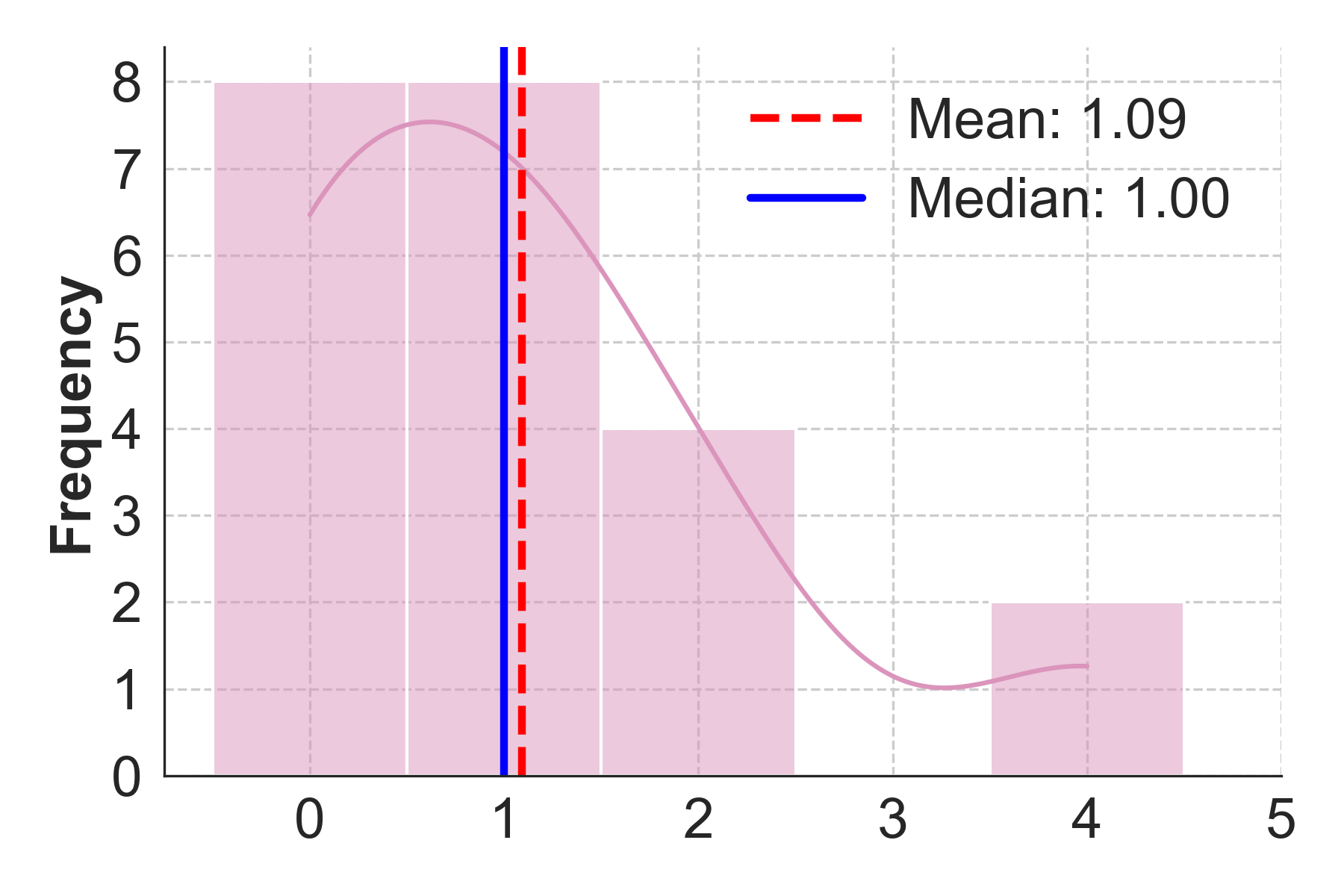}}  \subfloat[Cond B: Contrastive Questions Answered]{\includegraphics[width=4.1cm]{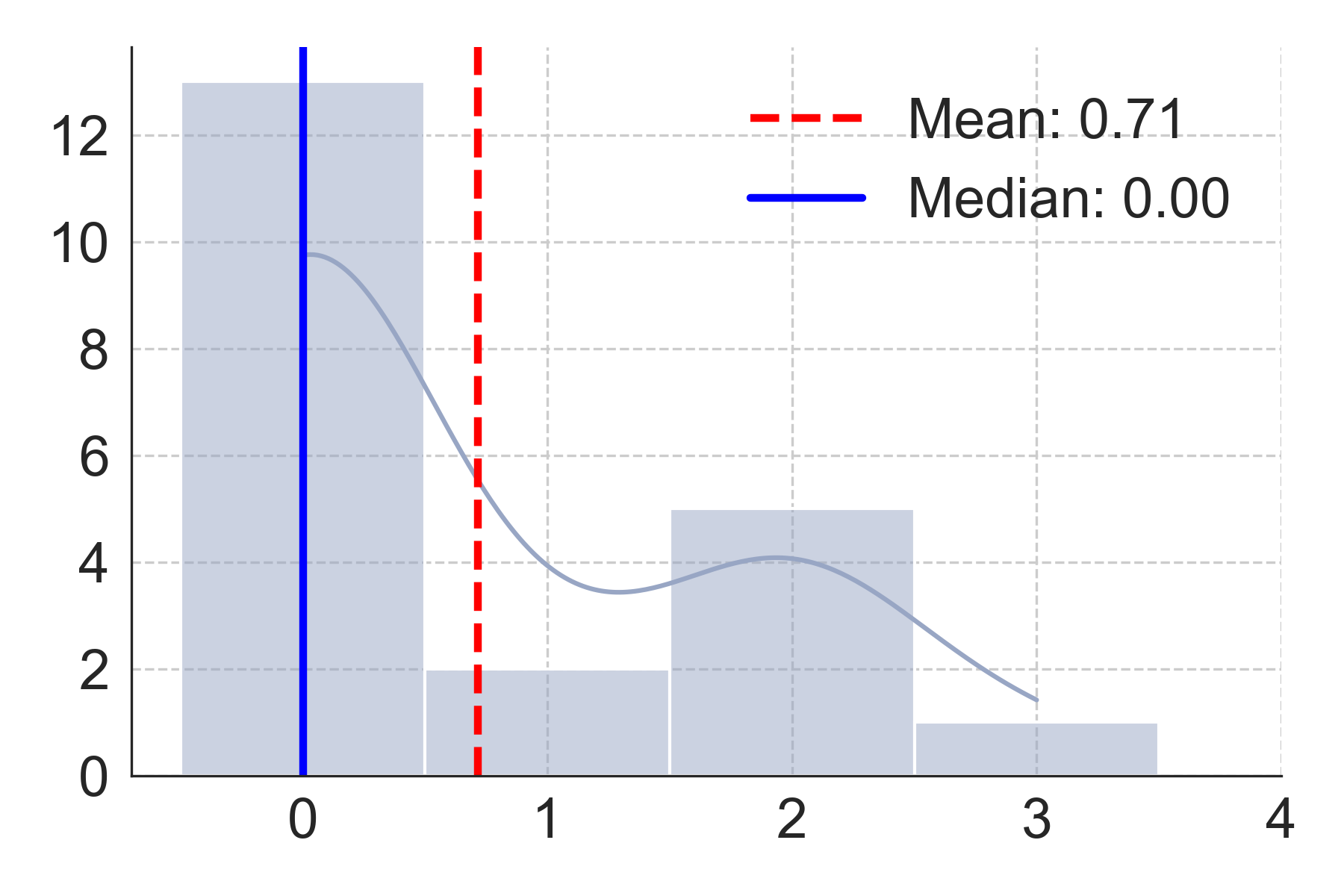}}\subfloat[Cond C: Hints Used]{\includegraphics[width=4.1cm]{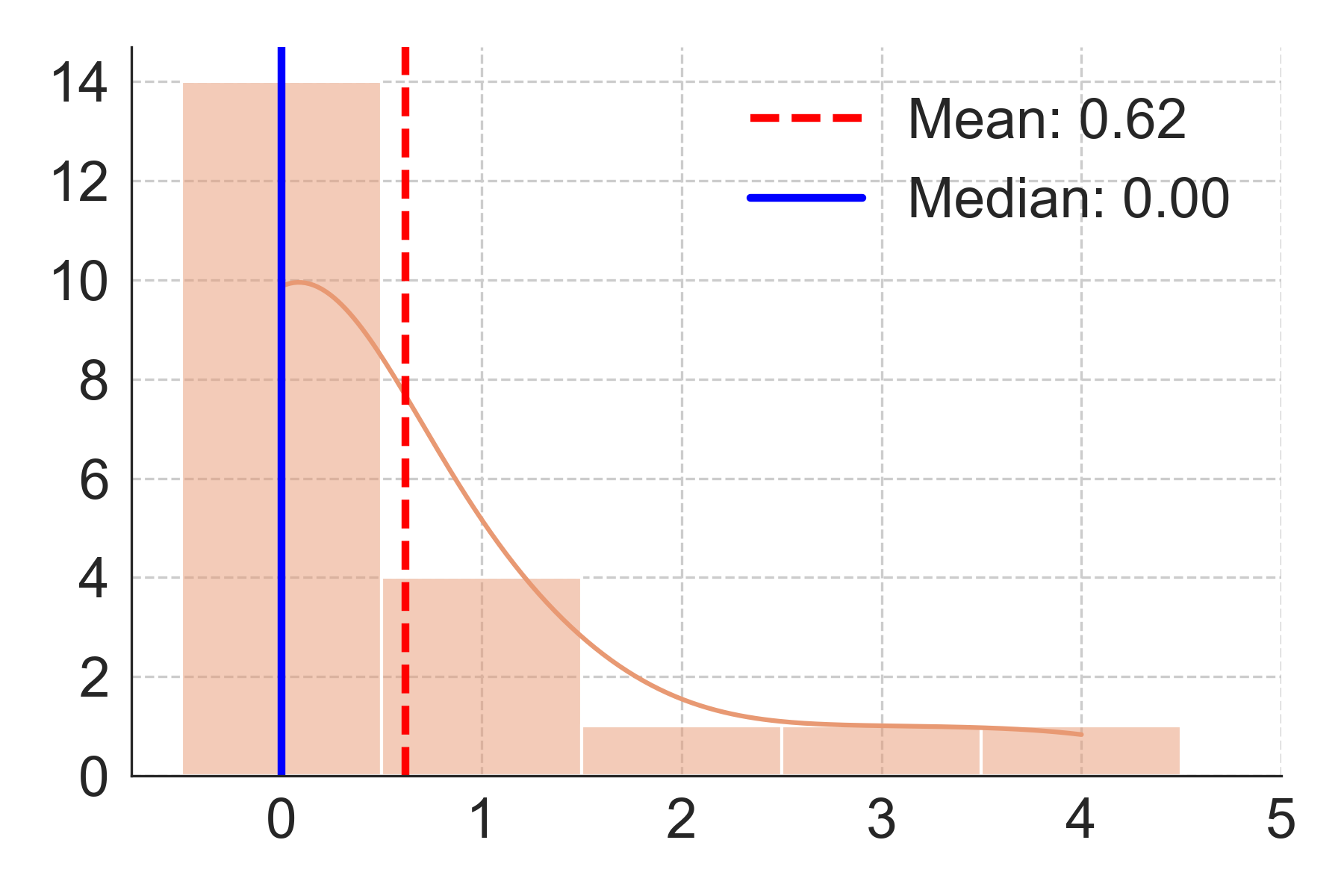}} \subfloat[Cond C: Solutions Used]{\includegraphics[width=4.1cm]{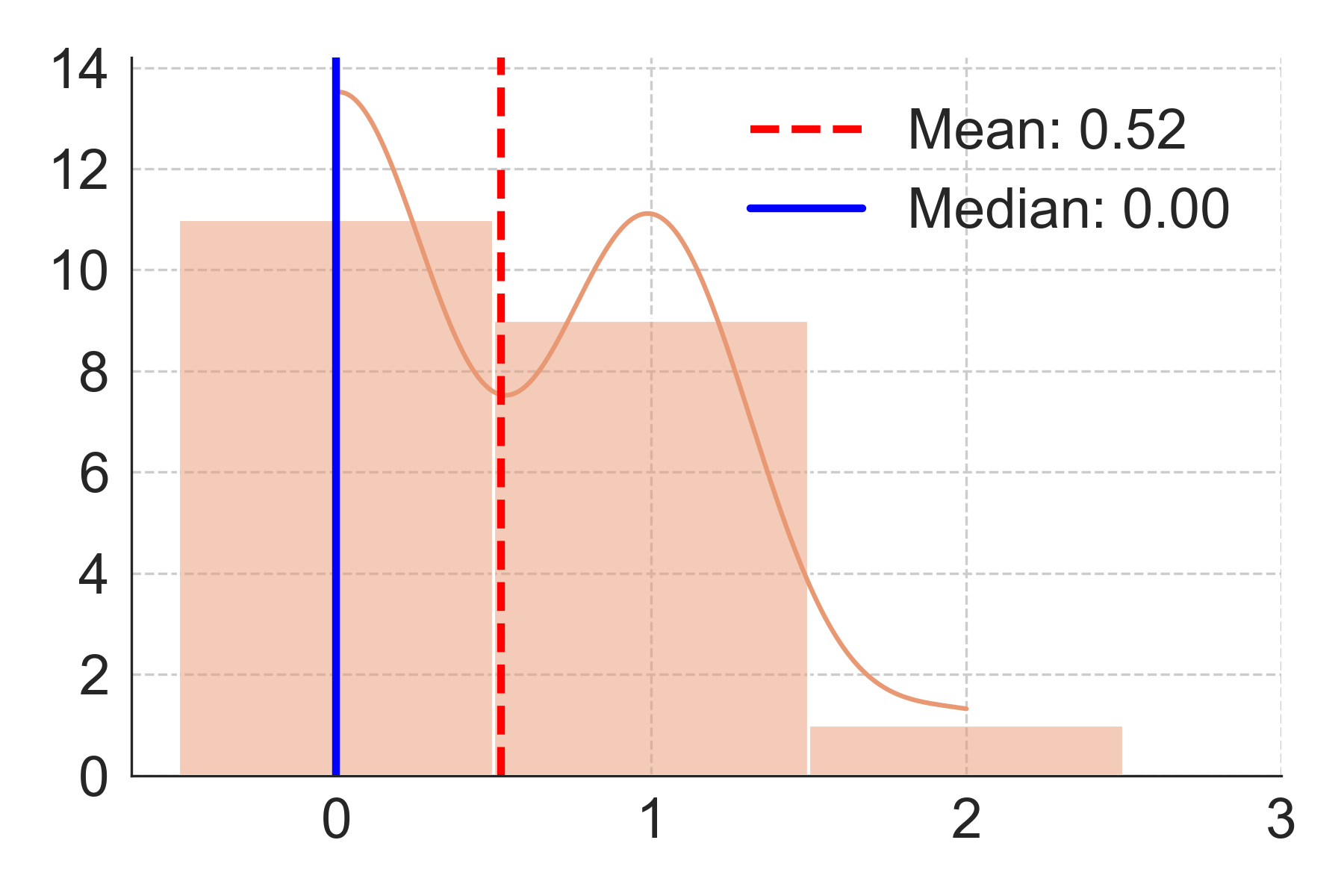}}
\caption{Task 2: Problem Solving. Condition specific feature usage across three interaction modes (Cond A: Future-Self Explanations, Cond B: Contrastive Learning, Cond C: Guided Hints). Histograms illustrate per-condition distributions across all participants; solid lines denote medians, dashed lines denote means, and the curve represents the kernel density estimation.}\label{conditionspecific2}
\end{figure*}

\begin{table*}[t!]
    \centering
    \caption{ Statistical summary of performance measures and tool usage across tasks. The "Overall" row for each measure reports the results of a one-way ANOVA ($F$) or Kruskal-Wallis ($H$) test. Subsequent rows indicate post-hoc pairwise comparisons between experimental conditions and the Baseline. Effect sizes are reported as Cohen’s $d$. Note. $\dagger p < .10, * p < .05, ** p < .01, n.s. = \text{not significant}$. }
    \label{tab:anova}
    \scalebox{0.9}{
    \begin{tabular}{lllll}
    \toprule
    \textbf{Task} & \textbf{Measure} & \textbf{Effect} & \textbf{Statistics} & \textbf{Sig.} \\ 
    \midrule
    \multirow{18}{*}{Task 1} & \multirow{4}{*}{Total Prompts} & Overall Effect & $H=4.61, p=0.203$ & $n.s.$ \\
    & & A vs Baseline & $t=2.28, p_{Bonf}=0.084, d=0.71$ & $\dagger$ \\
     & & B vs Baseline & $t=1.02, p_{Bonf}=0.951, d=0.32$ & $n.s.$ \\
     & & C vs Baseline & $t=1.00, p_{Bonf}=0.977, d=0.32$ & $n.s.$ \\
    \cmidrule{2-5}
     & \multirow{4}{*}{Completion Time (min)} & Overall Effect & $H=7.14, p=0.068$ & $\dagger$ \\
     & & A vs Baseline & $t=2.72, p_{Bonf}=0.030, d=0.85$ & $*$ \\
     & & B vs Baseline & $t=2.38, p_{Bonf}=0.071, d=0.75$ & $\dagger$ \\
     & & C vs Baseline & $t=1.22, p_{Bonf}=0.694, d=0.38$ & $n.s.$ \\
    \cmidrule{2-5}
     & Average Response Time & Overall Effect & $H=2.50, p=0.475$ & $n.s.$ \\
    \cmidrule{2-5}
     & Summaries Used & Overall Effect & $H=10.97, p=0.012$ & $*$ \\
    \cmidrule{2-5}
     & Quizzes Generated & Overall Effect & $H=5.84, p=0.120$ & $n.s.$ \\
    \cmidrule{2-5}
     & \multirow{4}{*}{Graded Understanding} & Overall Effect & $H=5.22, p=0.156$ & $n.s.$ \\
     & & A vs Baseline & $t=2.35, p_{Bonf}=0.073, d=0.74$ & $\dagger$ \\
     & & B vs Baseline & $t=1.00, p_{Bonf}=0.970, d=0.32$ & $n.s.$ \\
     & & C vs Baseline & $t=1.77, p_{Bonf}=0.254, d=0.56$ & $n.s.$ \\
    \midrule
    \multirow{18}{*}{Task 2} & \multirow{4}{*}{Total Prompts} & Overall Effect & $H=11.14, p=0.011$ & $*$ \\
    & & A vs Baseline & $t=1.86, p_{Bonf}=0.213, d=0.58$ & $n.s.$ \\
     & & B vs Baseline & $t=1.37, p_{Bonf}=0.543, d=0.43$ & $n.s.$ \\
     & & C vs Baseline & $t=3.88, p_{Bonf}=0.002, d=1.23$ & $**$ \\
    \cmidrule{2-5}
     & \multirow{4}{*}{Completion Time (min)} & Overall Effect & $F=2.49, p=0.066, \eta^2_p=0.084$ & $\dagger$ \\
      & & A vs Baseline & $t=2.41, p_{Bonf}=0.061, d=0.75$ & $\dagger$ \\
     & & B vs Baseline & $t=2.11, p_{Bonf}=0.124, d=0.67$ & $n.s.$ \\
     & & C vs Baseline & $t=2.32, p_{Bonf}=0.078, d=0.73$ & $\dagger$ \\
    \cmidrule{2-5}
     & Average Response Time & Overall Effect & $H=3.84, p=0.279$ & $n.s.$ \\
    \cmidrule{2-5}
     & Summaries Used & Overall Effect & $H=12.39, p=0.006$ & $**$ \\
    \cmidrule{2-5}
     & Quizzes Generated & Overall Effect & $H=5.86, p=0.119$ & $n.s.$ \\
    \cmidrule{2-5}
     & \multirow{4}{*}{Graded Understanding} & Overall Effect & $H=10.37, p=0.016$ & $*$ \\
     & & A vs Baseline & $t=2.22, p_{Bonf}=0.097, d=0.70$ & $\dagger$ \\
     & & B vs Baseline & $t=2.31, p_{Bonf}=0.080, d=0.74$ & $\dagger$ \\
     & & C vs Baseline & $t=3.64, p_{Bonf}=0.002, d=1.14$ & $**$ \\
    \bottomrule
    \end{tabular}
    }
\end{table*}

\begin{table*}[t!]
    \centering
    \caption{Statistical summary of questionnaire measures across tasks. The "Overall" rows report the results of a Mixed ANOVA ($F$) or Kruskal-Wallis ($H$) test. Subsequent rows indicate post-hoc pairwise comparisons between experimental conditions and the Baseline. Effect sizes are reported as Cohen’s $d$ or partial eta-squared ($\eta^2_p$). Note. $\dagger p < .10, * p < .05, ** p < .01, *** p < .001, n.s. = \text{not significant}$.}
    \label{tab:anova2}
    \scalebox{1}{
    \begin{tabular}{lllll}
    \toprule
    \textbf{Measure} & \textbf{Task} & \textbf{Effect} & \textbf{Statistics} & \textbf{Sig.} \\ 
    \midrule
    \multirow{10}{*}{Understanding} &  & Condition & $F=4.87, p=0.004, \eta^2_p=0.153$ & $**$ \\
     & & Task & $F=0.05, p=0.819, \eta^2_p=0.001$ & $n.s.$ \\
     & & Interaction & $F=1.40, p=0.250, \eta^2_p=0.049$ & $n.s.$ \\
     \cmidrule{2-5}
     &\multirow{3}{*}{Task 1} &  A vs Baseline & $t=2.35, p_{Bonf}=0.073, d=0.74$ & $\dagger$ \\
     & & B vs Baseline & $t=1.14, p_{Bonf}=0.778, d=0.36$ & $n.s.$ \\
     & & C vs Baseline & $t=1.77, p_{Bonf}=0.254, d=0.56$ & $n.s.$ \\
     \cmidrule{2-5}
     &\multirow{3}{*}{Task 2} & A vs Baseline & $t=2.22, p_{Bonf}=0.097, d=0.70$ & $\dagger$ \\
     & & B vs Baseline & $t=2.44, p_{Bonf}=0.057, d=0.77$ & $\dagger$ \\
     & & C vs Baseline & $t=3.64, p_{Bonf}=0.002, d=1.14$ & $**$ \\
    \midrule
    \multirow{3}{*}{NASA-TLX} & & Condition & $F=0.55, p=0.647, \eta^2_p=0.020$ & $n.s.$ \\
     & & Task & $F=63.58, p=0.000, \eta^2_p=0.443$ & $***$ \\
     & & Interaction & $F=0.71, p=0.551, \eta^2_p=0.026$ & $n.s.$ \\
    \midrule
    Trust & & Overall Effect & $H=0.64, p=0.886$ & $n.s.$ \\
    \midrule
    UEQ &  & Overall Effect & $F=1.08, p=0.362, \eta^2_p=0.039$ & $n.s.$ \\
    \bottomrule
    \end{tabular}
    }
\end{table*}

\begin{figure*}
\centering\subfloat[Perceived Task Understanding]{\includegraphics[width=4.2cm]{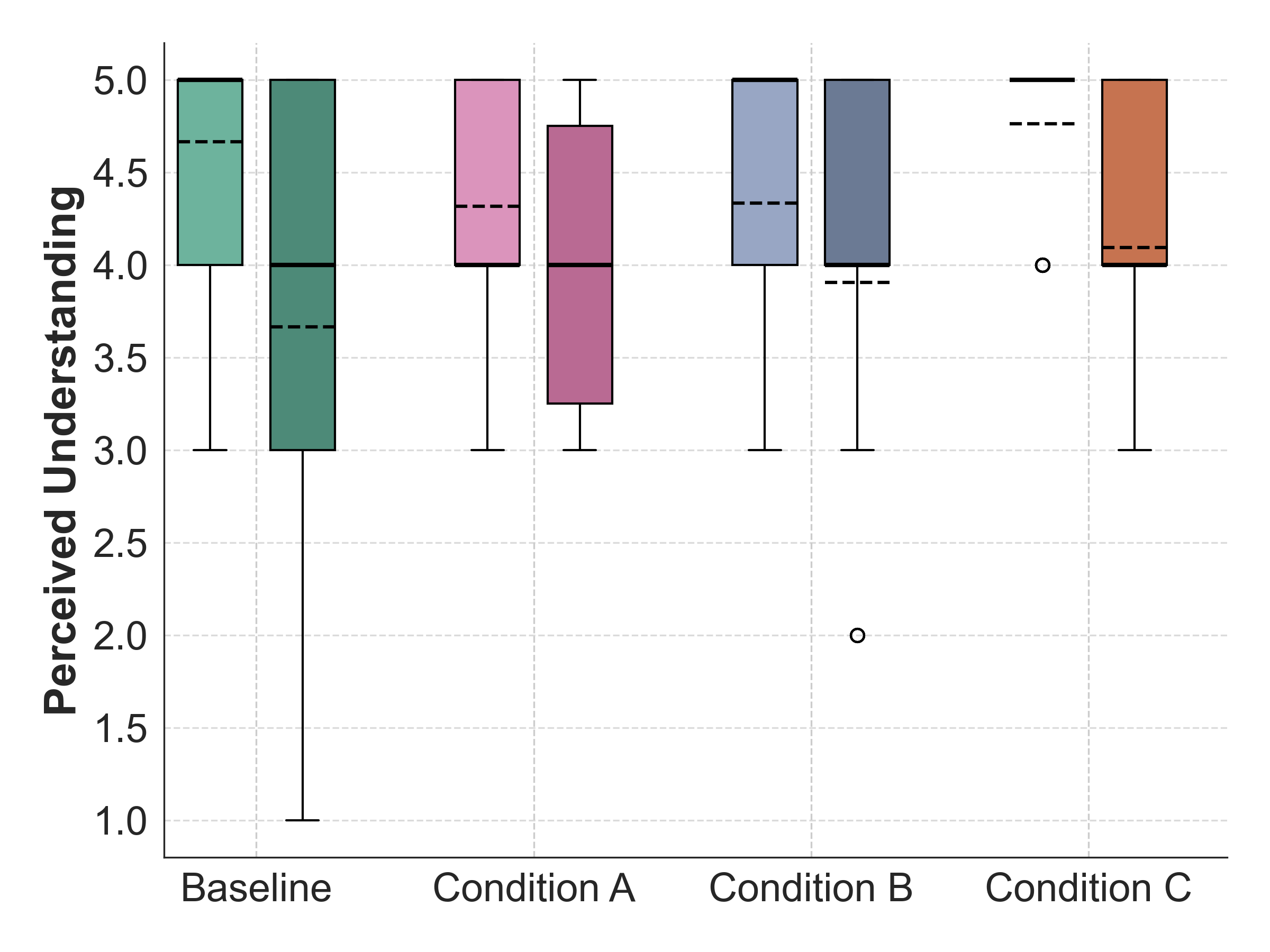}} \subfloat[NASA-TLX]{\includegraphics[width=4.2cm]{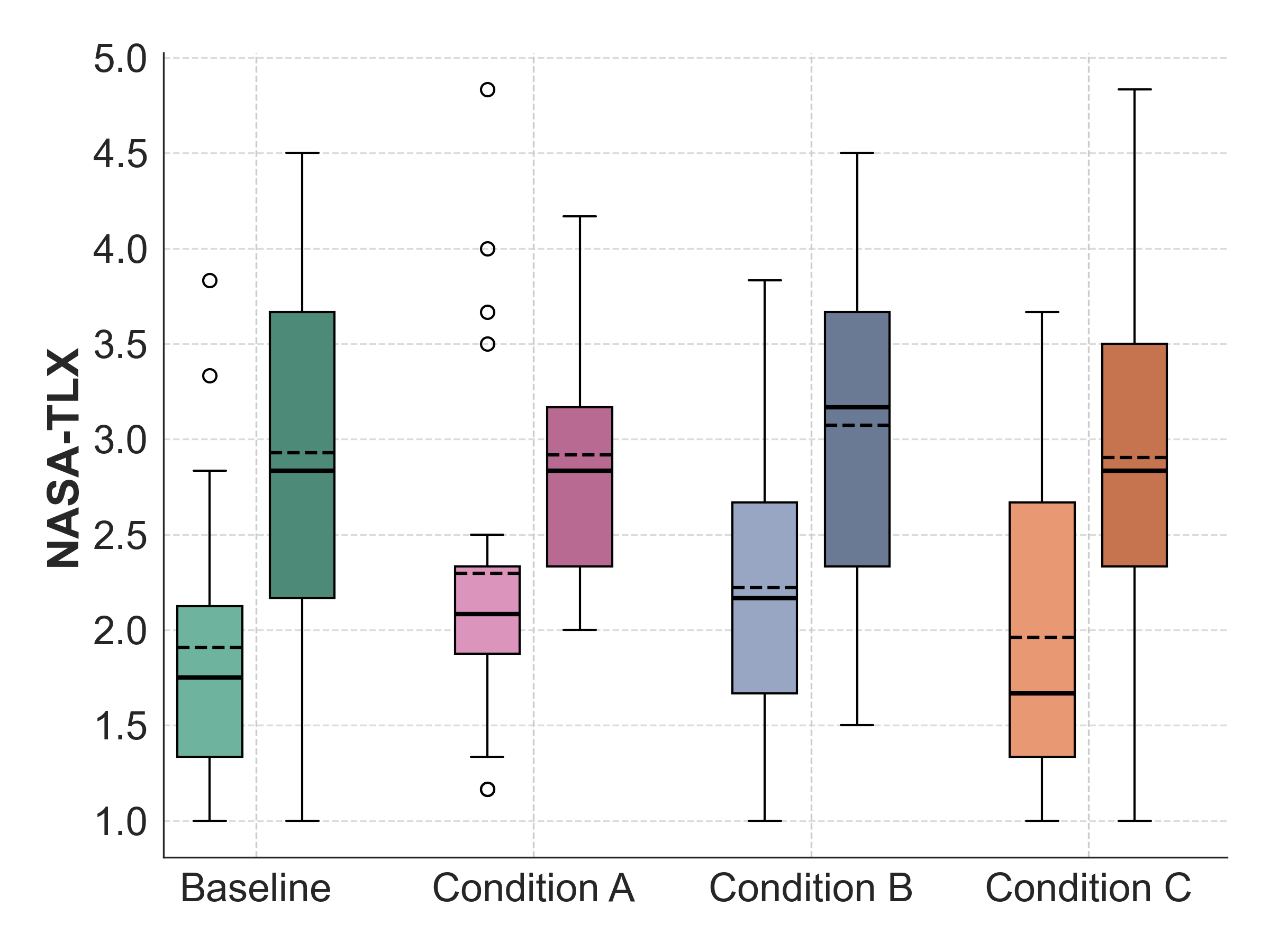}}\subfloat[Trust]{\includegraphics[width=4.2cm]{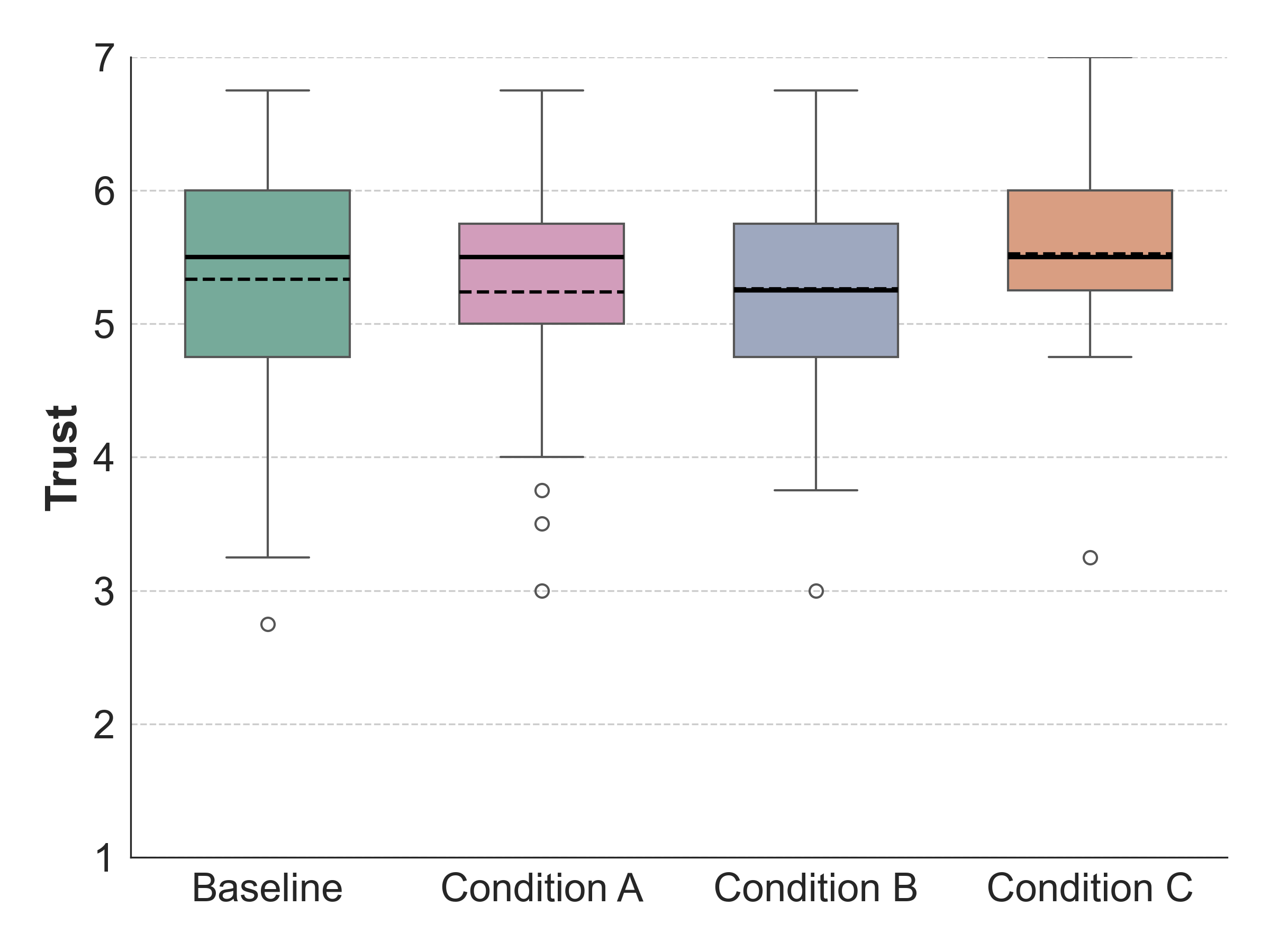}} \subfloat[User Experience]{\includegraphics[width=4.2cm]{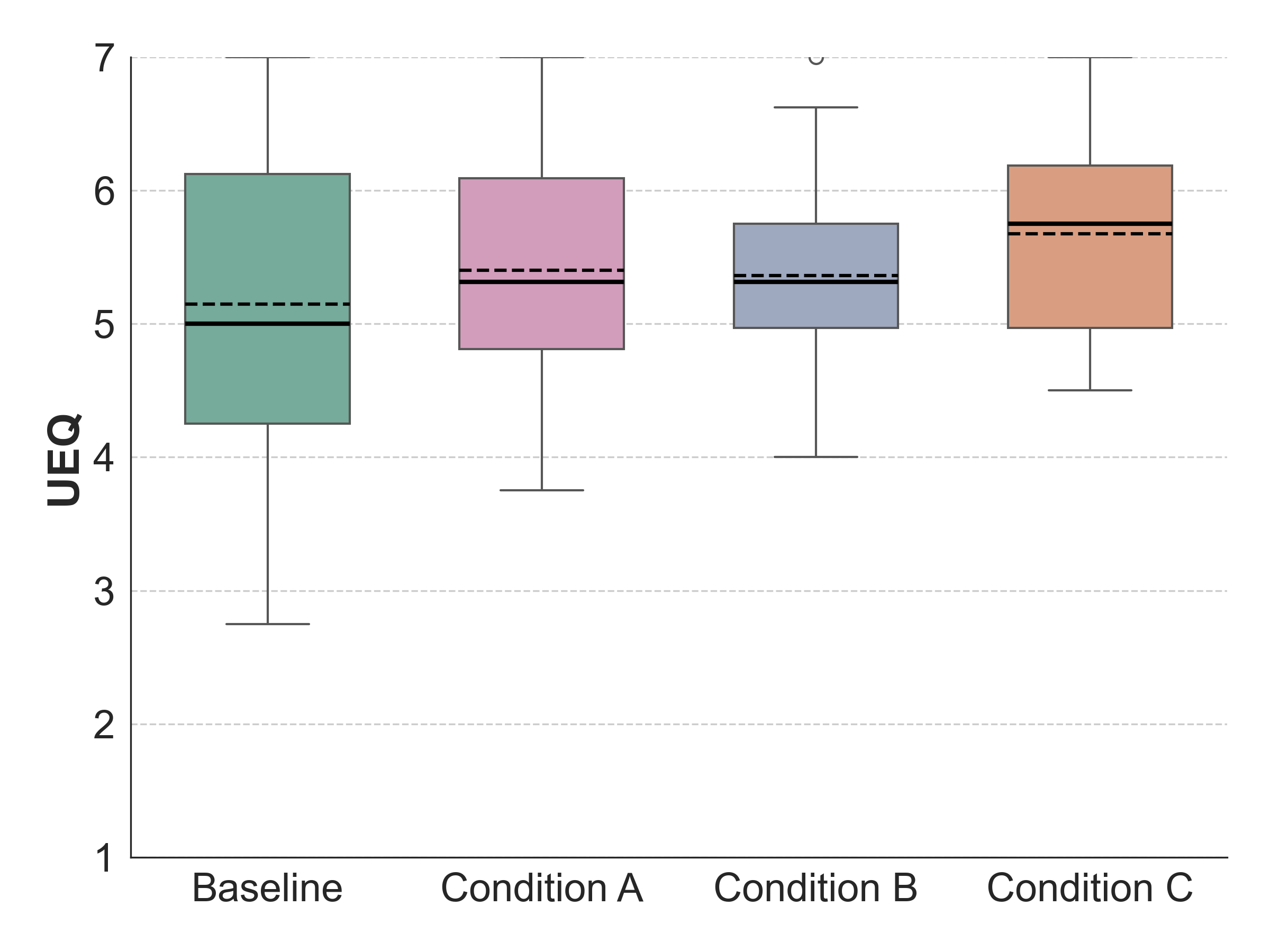}}
\caption{Boxplots of per-participant average questionnaire scores for the four interaction modes-Baseline, Cond A: Future-Self Explanations, Cond B: Contrastive Learning, Cond C: Guided Hints. Results are shown for both Task 1 (left) and Task 2 (right) for Perceived Task Understanding and NASA-TLX, and aggregated for Trust and User Experience. Solid lines denote medians; dashed lines denote means; whiskers indicate $1.5 \times \text{IQR}$; points mark outliers.}\label{questionnaires}
\end{figure*}

\begin{figure*}
    \centering
    \includegraphics[width=1\linewidth]{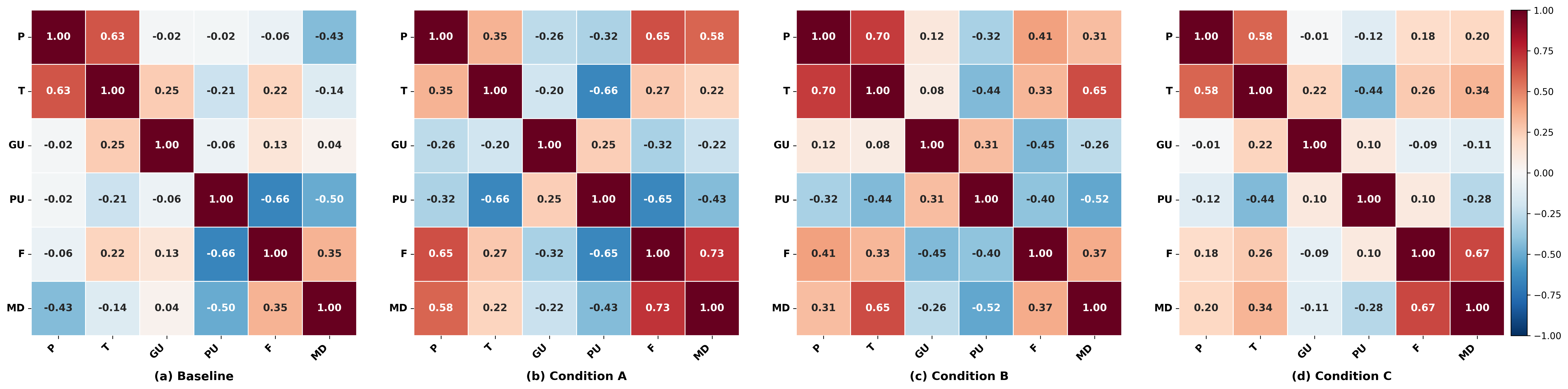}
    \caption{Cross-correlation matrices of four task-related metrics under each interaction condition in Task 1. The matrices show the Pearson correlation coefficients among \textbf{Prompts (P), Time (T), Graded Understanding (GU), Perceived Understanding (PU), Frustration (F) and Mental Demand (MD)}
    (ordered from top-left to bottom-right) for the four conditions: (a) Baseline, (b) Cond A: Future-Self Explanations, (c) Cond B: Contrastive Learning and (d) Cond C: Guided Hints. The color scale represents Pearson correlation coefficients ranging from $-1.0$ (blue, strong negative) to $+1.0$ (red, strong positive).}
    \label{fig:correlation1}
\end{figure*}

\begin{figure*}
    \centering
    \includegraphics[width=1\linewidth]{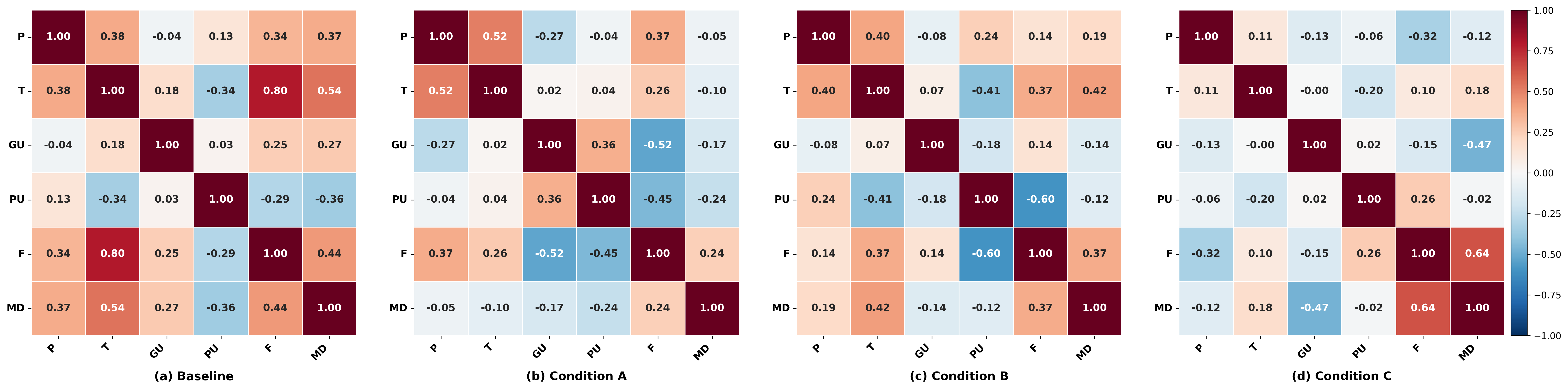}
    \caption{Cross-correlation matrices of four task-related metrics under each interaction condition in Task 2. The matrices show the Pearson correlation coefficients among \textbf{Prompts (P), Time (T), Graded Understanding (GU), Perceived Understanding (PU), Frustration (F) and Mental Demand (MD)}
    (ordered from top-left to bottom-right) for the four conditions: (a) Baseline, (b) Cond A: Future-Self Explanations, (c) Cond B: Contrastive Learning and (d) Cond C: Guided Hints. The color scale represents Pearson correlation coefficients ranging from $-1.0$ (blue, strong negative) to $+1.0$ (red, strong positive).}
    \label{fig:correlation2}
\end{figure*}

\subsection{Quantitative Analysis}

The distinction between the two experimental tasks is crucial for interpreting our findings. Our data reveals that Task 2 (Problem Solving) acted as a ``stress test'' that enhanced the differences between conditions~\cref{task2}. This task was significantly more demanding across all metrics, showing strong correlations between the overall workload and both total time ($r = 0.67$) and mental demand ($r = 0.49$). While NASA-TLX workload scores for Task 1 (Concept Explanation) were relatively low (1.93 to 2.38), Task 2 consistently triggered much higher scores (2.90 to 3.02).

The Baseline condition, which serves as our comparison point, clearly illustrates the difficulty of learning without AI support. This group exhibited a significant overconfidence effect. Participants reported high perceived understanding ($M = 4.65$ in Task 1) but achieved the lowest actual grades in the study ($M = 1.90$ for Task 1; $M = 1.40$ for Task 2)(~\cref{task1,task2}). This created a ``failure curve'' where, as the tasks became harder, the students' experience collapsed. While they stayed calm in Task 1, Task 2 showed an extremely strong correlation between the time they spent and the frustration they felt ($r = 0.80$)~\cref{fig:correlation2}. 

In contrast to the overconfident Baseline group, Condition A (Future-Self Explanations) acted as a model of efficiency for theoretical learning, with participants completing Task 1 significantly faster than the baseline ($d = 0.85, p = 0.030$)~\cref{tab:anova}. Even though the process was fast, it came with a ``high-friction'' experience. Users reported higher mental workloads and a strong link between their prompt usage and frustration ($r = 0.65$), leading to an ``underconfidence effect''. Another thing to note is that participants rated their own understanding at just $M = 4.32$ (on a 5-point scale) \revision{(Figure~\ref{questionnaires})}, which is the lowest self-rating among all experimental conditions, even though their actual test performance was quite high ($M = 2.36$ on a 3-point scale). Ultimately, Condition A led to the most ``calibrated'' understanding, meaning the students' perception of their knowledge closely matched their real performance. Finally, tool engagement remained stable, with participants viewing an average of $1.00-1.11$ self-explanations per task (~\cref{conditionspecific1,conditionspecific2}).

While Condition A focused on efficiency, Condition B (Contrastive Learning) served as a middle ground that balanced high cognitive load with a positive user experience. This condition was a distinct outlier, since it had the highest mental demand in the study ($M = 4.85$), yet participants maintained a high trust ($M = 5.24$) and the lowest reported frustration ($M = 2.00$). This suggests that the contrastive quizzes turned hard work into ``meaningful deliberation'', where the effort was perceived as useful rather than exhausting. System logs support this through a pattern of selective engagement, showing that participants were presented with roughly 2.5 quizzes per task but chose to answer only about 0.8 (~\cref{conditionspecific1,conditionspecific2}).  Ultimately, Condition B proved that an AI system can demand high cognitive effort without causing user frustration, if the struggle feels purposeful.

Condition C (Guided Hints) emerged as the most robust model, creating a seamless learning pathway that combined high engagement with the best academic results. In the second task (Problem Solving), this condition achieved a massive effect size for \revision{task performance} ($d = 1.14, p = 0.002$) and triggered the highest level of interaction, with a significant increase in total prompts ($d = 1.23, p = 0.002$). Remarkably, while increased interaction usually leads to higher stress~\cite{tarafdar2007impact}, Condition C demonstrated a ``frustration reversal'' ($r = -0.32$) ~\cref{fig:correlation2}. This means that as participants engaged more with the AI, their frustration actually decreased. This ``Invisible Scaffolding'' allowed students to achieve a 45\% improvement in learning outcomes over the baseline without any increase in perceived workload ($p = 0.647$). In fact, Task 2 workload scores for Condition C ($2.90$) were the lowest in the study. Finally, utilization of specific ``hint'' and ``solution'' buttons remained low ($Mdn = 0.00$, ~\cref{conditionspecific1,conditionspecific2}), suggesting that conversational support was sufficient to guide users without extra mechanisms.

To conclude our quantitative analysis, we examined the engagement patterns with features shared across all three experimental conditions: summaries and quizzes. Our analysis revealed a significant global effect for summary usage across both tasks (Task 1: $p = 0.012$; Task 2: $p = 0.006$). Interestingly, there were no significant differences in usage between Conditions A, B, or C (all $p_{Bonf} = 1.000$). Similarly, quiz generation remained consistent across all groups and tasks ($p = 0.119$). These findings suggest that while these tools were popular, the specific design of the AI agent did not change how often users clicked these buttons. 

\subsection{Qualitative Results}

By thematically analyzing our participants’ interviews, we identified three main themes: \textit{Cognitive Depth via Active Recall}, \textit{Perceived Knowledge Retention}, and \textit{Pedagogical Utility}. Each theme integrates qualitative insights from participant feedback with quantitative results where applicable, allowing us to see how the specific design and behavior of each AI agent shaped the way students interacted with the system and how they envisioned using these tools in their own studies. 
\smallskip

\noindent\textbf{Cognitive Depth via Active Recall.} The qualitative data confirms that while the Baseline was prefered for its low cognitive effort, it fostered a ``Fragile Understanding'' that collapsed under the stress of Task 2. Participants like P56 admitted the Baseline \textit{``simplifies things so it goes faster... [you] don’t use that much brain cells''}, directly explaining the high overconfidence and low \revision{task performance }($M=1.40$) seen in our quantitative analysis. In contrast, the experimental conditions, particularly Condition A, transformed the AI into a self-diagnostic tool. P30 noted, \textit{``I had to check just to be sure if I understood it right... I scrolled a bit up to see the actual answer''}, illustrating the ``productive friction'' that drove higher response times and mental demand. This shift toward ``desirable difficulty'' was also seen in Condition B, where P13 found that the AI \textit{``made me think more deeply... about points of view that maybe you hadn't considered''}, justifying the high mental demand ($M=4.85$) as a way to think deeply instead of just feeling frustrated. By moving away from the Baseline's \textit{``too many details''} (P14), the AI’s ability to work with the user’s existing logic allowed for the high-performing ``Invisible Scaffolding'' observed in Condition C. As P23 noted, \textit{``bridging that gap instead of just retelling me the definition''}. Ultimately, these features acted as vital \textit{``self-evaluation checkpoints''} (P8) that aligned the students’ understanding and ensured they were engaged through active thinking rather than passive consumption.
\smallskip

\noindent\textbf{Perceived Knowledge Retention.} The qualitative feedback highlights a divide between the ``Fragile Understanding'' of the Baseline and the thinking patterns frameworks fostered by the AI interventions. Participants in the Baseline group admitted a ``shallow'' experience, for example when asked if the tool encouraged deeper thought, P19 briefly replied, \textit{``No''}. This explains the low NASA-TLX workload in Task 1 \revision{(Figure~\ref{questionnaires})} but also the subsequent collapse in \revision{task performance} during Task 2. When faced with complex problems, Baseline users like P14 found the AI's detailed answers ``annoying'' and felt they were simply ``too many details'' to process, leading to a failure in fully understanding the idea. In contrast, participants in the experimental conditions linked the effortful interaction to long-term memory. P30 (Cond A) explicitly stated, \textit{``You remember what you understand. So if I have to explain it, I first must understand''}, suggesting that slow learning and high mental effort were necessary for long-term memory. This was further refined in Condition C, which users perceived as a \textit{``step-by-step tutorial on how to evolve your thinking process''} (P23). P32 noted that this scaffolded support became a repeatable mental framework, predicting they could \textit{``think back to your process with the bot'' }during an exam to apply the same logic. This qualitative evidence clarifies why Condition C achieved a massive effect size in \revision{task performance} ($d = 1.14$), it didn't just provide answers, it helped users internalize the logic required to find them. While some users found the contrastive approach in Condition B ``complicated'' for strict definitions (P50), the shared view across experimental groups was that features like quizzes served as necessary ``self-evaluation'' checkpoints (P8, P23) to help their progress. 
\smallskip

\noindent\textbf{Pedagogical Utility.} The qualitative feedback underscores a fundamental shift in the AI’s role from a simple information provider to a sophisticated pedagogical agent. While the Baseline was initially seen as efficient, it ultimately failed as a teaching tool. Participants like P14 and P19 noted it never prompted deeper thought, leading to a ``slower'', more frustrating experience when tasks became complex. In contrast, the experimental conditions were perceived as high-level learning strategies. Condition B, in particular, was recognized for its academic rigor, with P17 noting the system \textit{``had a way of teaching like professors''}, which directly explains the high trust ratings ($M = 5.24$). Furthermore, participants noted that while Condition B is ideal for theoretical fields, it may require more \textit{``strict definitions''} (P50) for applied sciences. Condition C achieved a balance of ``Invisible Scaffolding'', as P32 expressed a desire to use the tool in a university setting because \textit{``it helps me think more rather than just giving me an answer''}. This sense of partnership fostered the ``frustration reversal'' ($r = -0.32$) seen in the logs, as users felt the system was \textit{``bridging the gap''} (P23) rather than just lecturing. Condition A functioned as a mechanism for metacognitive verification and gap identification through its "explain-back" prompts. This utility was validated by P30, who noted it automated a strategy they already use in university: \textit{``I ask something and then... I always say I understood this. Is it right?''} Noting the need for a tool that automates this process.
\section{Discussion}
\label{sec:discussion}

\begin{table*}[t!]
    \centering
    \caption{Summary of qualitative and quantitative findings across interaction modes.}
    \label{tab:findings_summary}
    \setlength{\tabcolsep}{4pt} 
    \scalebox{0.8}{ 
    \begin{tabular}{lp{3.2cm}p{3.2cm}p{3.2cm}p{3.2cm}}
    \toprule
    \textbf{Dimension} & \textbf{Baseline} & \textbf{Condition A} & \textbf{Condition B} & \textbf{Condition C} \\ 
    \midrule
    Performance & Lowest grades ($M=1.40$ in T2). & High accuracy ($M=2.36$); sig. faster T1 ($d=0.85^*$). & Sig. improvement over baseline (+1.00 pt). & \textbf{Best results}; massive effect size in T2 ($d=1.14^{**}$). \\
    \addlinespace
    Metacognition & \textbf{Overconfidence}: High perceived vs. low actual knowledge. & \textbf{Metacognitive Anchor}: Lowest self-rating ($M=4.32$) despite high test scores. & Balanced; perceived as ``Meaningful Deliberation''. & High calibration; logic internalized as mental framework. \\
    \addlinespace
    Cognitive Load & Low initial effort; ``Fragile Understanding'' in T2. & High mental demand; perceived as ``High-Friction''. & \textbf{Highest mental demand} ($M=4.85$); effort separated from frustration. & Lowest T2 workload ($2.90$); ``Invisible Scaffolding''. \\
    \addlinespace
    Emotional Response & Time-Frustration correlation ($r=0.80$); ``Failure Curve''. & Prompt usage linked to frustration ($r=0.65$). & \textbf{Lowest frustration} ($M=2.00$); high trust ($M=5.24$). & \textbf{Frustration Reversal}: Interaction reduced stress ($r=-0.32$). \\
    \addlinespace
    Tool Engagement & Passive; AI answers labeled ``too many details''. & Stable feature usage; $\approx 1.1$ explanations/task. & Selective engagement; quizzes answered only when relevant. & Highest interaction; sig. increase in prompts ($d=1.23^{**}$). \\
    \addlinespace
    UI Interaction & n/a & Primary use of ``Explain Back'' prompts. & Quizzes used as ``Self-evaluation checkpoints''. & \textbf{External buttons ignored}; conversational foundation was enough. \\
    \bottomrule
    \end{tabular}
    }
\end{table*}

Next, we situate our findings within prior work on GenAI in education, productive friction and active learning. We first highlight our main findings, drawing from both quantitative and qualitative data. We then discuss the implications of our results, outline contributions to HCI theory and design practice, as well as its ethical considerations before acknowledging the limitations of our study and identifying directions for future research.

\revision{It is important to consider the volume of information provided across conditions. While the baseline condition provided a single, comprehensive answer, consistent with standard LLM behaviors like ChatGPT, the experimental conditions were intentionally designed to be more concise, typically limiting output to 1–2 paragraphs. Additional features, such as guided hints, were restricted to 1–2 sentences to avoid overwhelming the user. Consequently, the experimental conditions provided less immediate information than the baseline. Also, the Presentation themes (Section~\ref{presentation_themes}), were implemented only to the experimental conditions, as the Baseline was designed to mirror standard current practices without additional interventions, providing a representative control for existing LLM interaction models.}

\subsection{Main Findings}

Our results show that the way AI scaffolding is designed changes how LLMs are used in education (i.e., turning them from simple information sources into tools for thoughtful learning). By using two different tasks, we showed how different teaching methods balance effort and understanding. Our findings reveal four key insights: 1) Condition A provides a metacognitive reference point, aligning \revision{task understanding with task performance}; 2) Condition B creates a ``desirable difficulty'', turning high mental effort into meaningful thinking without causing extra frustration; 3) Condition C provides subtle support, helping users absorb logical frameworks and reach top problem-solving performance without asking for hints; and 4) there is no "one-size-fits-all" AI tutor, rather, the optimal intervention depends on the learner's priority.
\smallskip

\noindent\textit{Condition A provides a metacognitive reference point, aligning \revision{task understanding with task performance.}} By requiring users to explain concepts back to the AI, this condition acted as a metacognitive anchor, forcing participants to confront gaps in their own logic~\cite{chi1989self}. While the baseline group reported high \revision{task understanding }($M=4.65$) despite having the lowest objective scores ($M=1.40$), Condition A reversed this trend. Our data highlights the fact that participants achieved the study’s highest accuracy ($M = 2.36/3$) while rating their own comprehension lower than all other groups ($M = 4.32/5$). This suggests that the "Correction Loop" works as a self-diagnostic tool, as P30 noted, they were forced to \textit{``check just to be sure''} if they understood. This early challenge, shown by the Task 1 frustration correlation ($r=0.65$), ultimately led to the most calibrated relationship between perceived and actual knowledge.
\smallskip

\noindent\textit{Condition B creates a ``desirable difficulty'', turning high mental effort into meaningful thinking without causing extra frustration.} It caused the study’s highest mean mental demand ($M=4.85$), yet it effectively separated effort from frustration. While Task 2 was a ``stress test'' for other conditions, Condition B maintained strong trust ratings ($M=5.24$) and the study's lowest reported frustration ($M = 2.00$). The qualitative feedback suggests this was achieved through learner autonomy. Users were selective in their engagement with the counterarguments, answering only contrastive quizzes they deemed relevant~\cite{buccinca2025contrastive}. This allowed them to treat the AI like a \textit{``professor''} (P17)~\cite{danry2023don}, engaging in ``meaningful deliberation''. By offering contrastive foils, the system prompted users to think \textit{``more deeply''} (P13) about alternative points of view, resulting in a significant 1.00-point improvement in objective scores over the baseline without a significant increase in frustration.
\smallskip

\noindent\textit{Condition C provides subtle support, helping users absorb logical frameworks and reach top problem-solving performance without asking for hints.} This condition achieved the most robust quantitative result in the study, with a big effect size for \revision{task performance} ($d=1.14$), while simultaneously reporting the study's lowest cognitive workload in Task 2 ($2.90$). The effectiveness of this intervention is best shown by the ``frustration reversal'' ($r = -0.32$). While the baseline showed a $r=0.80$ correlation between time and frustration, Condition C proved that more interaction with it actually reduced user stress. Remarkably, participants achieved these results while rarely using the ``hint'' or ``solution'' buttons \cref{conditionspecific2}. This indicates that the agent's guidance was so effective that users learned the AI’s logical approach, evolving their thinking process to a \textit{``step-by-step''} (P23) approach rather than relying on external cheat mechanisms~\cite{sarkar2024copilot}.
\smallskip

\noindent\textit{Ultimately, these results suggest that there is no "one-size-fits-all" AI tutor, rather, the optimal intervention depends on the learner's priority.} For learners seeking conceptual depth, Condition A is most effective. For those seeking meaningful deliberation and theoretical exploration, the contrastive challenges of Condition B provide a balanced academic rigor. For students requiring applied mastery and cognitive efficiency during complex problem-solving, the scaffolding of Condition C is the best approach. These findings suggest that future LLM-based tutors must move beyond being simple ``information providers''~\cite{danry2023don} to become adaptive systems~\cite{li2025tutorllm} capable of shifting their intervention style based on the task’s cognitive demand and the learner’s specific pedagogical goals.

\subsection{Implications}

Our findings demonstrate that cognitive load and user frustration are not directly linked. This challenges the traditional UX dogma that friction necessarily leads to a poor user experience~\cite{evans2024cognitive,hart2006nasa}. Although~\citet{kosmyna2025your} suggested that LLM dependence leads to lower brain engagement, our results for Condition B and C show that AI can facilitate ``Desirable Difficulty'' without triggering the emotional ``failure curve'' ($r = 0.80$) seen in the Baseline. Specifically, our ``frustration reversal'' finding suggests that when AI provides scaffolding rather than direct answers, the mental effort feels like meaningful exploration, not just an obstacle~\cite{park2023thinking}. This extends Cognitive Load Theory by showing that AI can reduce emotional strain, helping students stay in “Flow” even when tasks get harder.

A major practical takeaway for the design of Intelligent Tutoring Systems is the under utilization of explicit help mechanisms when compared to integrated dialogue. While our experimental interface provided participants in Condition C with dedicated ``Hint'' and ``Solution'' buttons, these features were ignored ($Mdn = 0.00$) \cref{conditionspecific2}. This suggests that when an AI provides strong conversational foundation, users do not perceive a need for extra UI interventions. This finding advances the work of systems like Iris Tutor \cite{bassner2024iris} and TutorLLM \cite{li2025tutorllm} by demonstrating that effective scaffolding does not require separate UI triggers. By ``formatting'' the user’s existing logic rather than simply ``rephrasing'' it (P23), the AI bridges the gap between the student's current mental model and the target solution.  For designers, the implication is clear: instead of building explicit ``help'' buttons, LLMs should be prompted to reflect the user's own reasoning back to them. This ``Invisible Scaffolding'' allows users to internalize the logic as a repeatable mental framework, supporting the long-term skill gain highlighted by~\citet{rogers2025worth}.

Our results for Condition A offer a guide for closing the metacognitive gap in Gen Z’s AI usage~\cite{olander2025}. While~\citet{lee2025impact} noted that high confidence in AI often leads to less critical thinking, Condition A acted as a ``Metacognitive Anchor'', forcing users to confront what they did not know. The resulting ``Underconfidence Effect'' is a vital insight for the HCI community. It suggests that the best learning tools might lower short-term satisfaction by challenging users’ overconfidence in their understanding~\cite{amoah2025chatgpt}. Using AI in education requires, therefore, balancing accurate self-assessment with positive encouragement. This ensures that the initial cognitive friction observed in our correlation data ($r = 0.65$) does not lead to premature tool abandonment. 

Finally, our work presents a practical taxonomy of AI interventions, showing that the “best” agent depends on context. Condition A is ideal for Conceptual Grounding, using active recall to confirm understanding before students move to application. Condition B supports Meaningful Deliberation, especially in theoretical tasks that challenge students to consider multiple perspectives. Condition C works best for Problem-Solving exercises, helping students achieve applied mastery in domains where logical reasoning is crucial. This functional split suggests that future educational platforms should move toward adaptive, multi-agent ensembles, similar to the architecture in EduThink4AI~\cite{hou2025eduthink4ai}, that shift their persona.

\subsection{Ethical Considerations and Design Trade-offs}

The findings of this study raise ethical and design trade-offs regarding the balance between user autonomy and pedagogical enforcement. A primary tension emerged in Condition A, where the \textit{``Explain what you understood"} prompt was occasionally misinterpreted as part of the AI’s output rather than a mandatory user action. As noted by P8 and P24, the lack of explicit UI cues, such as a dedicated Explain button, led to confusion, leaving users unsure if participation was required. This points to a key design trade-off between maintaining smooth conversational flow in LLM interfaces and enforcing stricter UI rules to introduce intentional challenges. Designers must decide whether to prioritize a seamless experience or to implement mandatory steps that ensure participants apply the required mental effort for deep learning.

Similarly, Condition B revealed an ethical challenge regarding selective engagement and automation bias. While the ability to ignore contrastive foils and contrastive questions provided users with a sense of autonomy and reduced frustration, it also allowed them to bypass the very meaningful deliberation the system was designed to provoke. This mirrors broader concerns in HCI regarding automation bias, where users may prefer the most efficient path over the most demanding one. If users are permitted to skip cognitive challenges, the system risks regressing into a ``Baseline'' one. Designers must decide when to let users manage their own cognitive load and when to require engagement with counter-arguments to help them fully understand concepts.

Our results also suggest that the effectiveness of reflective AI depends on the subject area, raising questions about the universal applicability of these models. Participants like P50 expressed doubt regarding the utility of contrastive learning for applied sciences such as mathematics or physics, where strict definitions and singular logical paths are prioritized over the multi-perspective debate common in classical studies or literature. There is a risk that deploying a contrastive agent in a domain requiring high-precision problem-solving could introduce unnecessary cognitive interference that confuse the learner. This suggests an ethical responsibility to ensure that AI tutoring architectures are tailored to the specific requirements of the subject matter.

The Underconfidence Effect observed in Condition A introduces a psychological trade-off for long-term tool adoption. While the intervention successfully destroyed the ``Illusion of Competence'' and led to high objective performance, it also left users feeling less capable. In a commercial or optional educational setting, a tool that makes a user feel ``slower'' or ``less smart'' faces a significant risk of abandonment. Designers must therefore consider the ethical implications of corrective AI: how do we design systems that are honest about a student’s knowledge gaps without damaging their self-confidence? Balancing the “Correction Loop” with positive feedback or progress tracking may be key to preventing the early demands of deep learning from discouraging learners.

\revision{Finally, a key challenge raised is the ``adoption barrier'' that is part of introducing specialized educational tools in an ecosystem dominated by general-purpose LLMs like ChatGPT. Even though students who tried the tool liked it, many are unlikely to switch from the tools they already use. Instead of trying to replace those tools, developers should try to improve them. The goal should be to quietly add our features into the tools students already use, so it feels natural and easy to adopt.}

\subsection{Limitations and Future Work} 

In Phase 1, we acknowledge two primary limitations. First, participants were recruited through convenience sampling, which may limit the generalization of our findings to the broader student population. Second, our sample exhibited a high degree of AI literacy and prior exposure. Nearly all participants reported regular experience with Generative AI tools, with none identifying as non-users or AI-skeptics. Future research should specifically target AI-naive users or those with a skeptical stance toward automated tutoring to understand broader needs and adoption barriers.

In Phase 2, we acknowledge seven limitations. \revision{First, our recruitment strategy was relied on personal networks and institutional advertisements, which may have introduced social desirability bias, where participants who know the researchers may perform differently or rate the system more favorably. To address these concerns, future studies should utilize independent facilitators to ensure that participant feedback remains objective and uninfluenced by personal connections to the research team.} 

Second, our participant pool ($N=85$) was primarily comprised of undergraduate students from STEM disciplines, particularly Computer Science ($N=24$). Future work should aim for a more diverse sample to examine how these AI interventions perform across different academic fields and levels of expertise. 

\revision{A third limitation is about collecting demographic data at the start of the study may have accidentally caused participants to feel a stereotype threat, potentially biasing performance or self-reporting (NCWIT~\cite{ncwit_evaluation_tools}). Future work should relocate these questions to the end of the protocol to mitigate identity-based priming effects.} 

\revision{Fourth}, we did not examine long-term engagement and knowledge retention. Although approximately 15\% of participants reported a desire for more time to deeply engage with the agents, our measurements were limited to immediate performance gains. Consequently, we cannot determine whether the ``Underconfidence Effect'' in Condition A leads to tool abandonment over weeks of use, or whether the logic internalized in Condition C persists in the absence of the AI. Future work could investigate the integration of these three conditions into existing learning systems to evaluate their long-term efficacy. 

\revision{Fifth, the study is also limited by its exclusive reliance on student self-reports. Literature suggests that students are often ``unreliable narrators'' of their own learning, frequently mistaking engagement for actual cognitive gain \cite{bowman2010can}. Consequently, future work should expand this inquiry to include educator perspectives, which remain highly varied regarding the role of AI in instruction \cite{lau2023ban}.} 

\revision{Sixth,} we utilized the llama-4-maverick-17b-128e-instruct model as the underlying LLM. While this model worked well for our study, future work should integrate newer ``reasoning-focused'' AI models because they might provide even better conversational support. 

\revision{Lastly}, although demographic data were collected, these variables were not included as controls in the statistical modeling of interaction logs. Our analytical choice prioritised a model that captures behavioural patterns emerging from the interaction designs themselves, rather than individual background characteristics. Consequently, we do not make claims about how factors such as field of study or prior LLM familiarity may moderate these effects. Future research should employ multi-variable modeling to isolate the effect of these individual differences. Ultimately, the goal is to refine these pedagogical agents so they don't just provide answers, but actively foster the ``meaningful deliberation'' necessary for Gen Z students to succeed in an AI-integrated academic landscape.
\section{Conclusion}

In this study, we examined how diverse interaction architectures shape the educational utility and user experience of LLM-based tutors. Our findings show that how an AI intervention is designed directly shapes the key features of the learning process, proving that the best AI agent is strictly context-dependent. Condition C emerged as best for problem-solving exercises, providing ``invisible scaffolding'' that achieved the study’s peak learning effect size without the need for explicit support. In contrast, Condition A functioned as a metacognitive anchor, neutralizing the ``illusion of competence'' observed in baseline interactions by forcing active recall. Condition B successfully transmuted high mental demand into ``meaningful deliberation'', demonstrating that contrastive foils can challenge a student's view without creating emotional burnout. Ultimately, these results demonstrate that while a ``Baseline'' interaction satisfies a user’s ``Speed Addiction'', it hides a lack of conceptual depth, but, by intentionally creating productive friction, we can move toward adaptive agents that empower genuine conceptual internalization. Together, these findings show how carefully designed interactions can reduce the risks of over-relying on generative AI and guide more effective human-AI collaboration in education.

\label{sec:conclusion}

\section{AI Disclosure Statement}
LLMs were used as the primary object of study and interaction in this research. The Deep3 system integrates an LLM to generate responses, prompts, counterarguments, and hints during participant tasks. Participants were informed that they were interacting with an AI-based system and consented to its use as part of the study.

LLMs were also used during the research process in limited and non-substantive ways. This included assistance with code debugging and language polishing of draft text. No AI systems were used to generate research ideas, formulate research questions, analyze qualitative or quantitative data, interpret results, or write analytic sections of the paper. All data analysis, interpretation, and final writing decisions were conducted by the authors.

\begin{acks}
We thank all study participants for their time and contributions. 

\end{acks}

\bibliographystyle{ACM-Reference-Format}
\bibliography{main}

\newpage
\appendix
 
\onecolumn
\section{\revision{Appendix}}

\subsection{Interview Protocol for the Formative Study} 
\label{app:interview_questions}

The following questions were used during the 20-30 minute semi-structured interviews with participants P1-P16. 

\subsection*{Phase A: Current LLM Usage and Critical Thinking}
\begin{enumerate}
    \item Can you describe how you currently use LLMs (e.g., ChatGPT, Copilot) in your studies?
    \begin{itemize}
        \item What do you usually ask it to do for you?
        \item Do you use it more for quick answers, or for longer projects?
    \end{itemize}

    \item Can you walk me through a recent task where you used an LLM? What steps did you take?
    \begin{itemize}
        \item What was the task about?
        \item Which part of the task did you use the LLM for?
        \item Did you try more than one prompt or just one?
    \end{itemize}

    \item How do you decide which outputs from the LLM to use or modify?
    \begin{itemize}
        \item What makes an answer feel ``good enough'' to you?
        \item Do you usually check the information somewhere else?
    \end{itemize}

    \item How much of your work is typically your own versus influenced or copied from the AI?
    \begin{itemize}
        \item Would you say it’s more like a first draft from you, or from the LLM?
        \item Do you often rewrite the AI’s words into your own style?
    \end{itemize}

    \item In your experience, does using LLMs help you think more critically or reflect on the task? Can you give an example?
    \begin{itemize}
        \item Has it ever made you notice something you hadn’t thought of before?
        \item Do you feel you understand the topic better after using it?
    \end{itemize}

    \item Are there moments when using an LLM makes it harder to engage deeply with the task?
    \begin{itemize}
        \item Do you ever feel like you rely on it too much?
        \item Does it sometimes make you skip doing your own reasoning?
    \end{itemize}
\end{enumerate}

\subsection*{Phase B: Design Needs and Preferences}
\begin{enumerate}
    \item If we were designing a tool to help you think more critically while using LLMs, what features would be most helpful?
    \begin{itemize}
        \item What would make it easy to use?
        \item Would you prefer text, visuals, or something interactive?
        \item Can you think of one feature you’d definitely want?
    \end{itemize}

    \item How could a tool encourage you to reflect on LLM suggestions rather than just accept them?
    \begin{itemize}
        \item What would make you stop and think for a moment?
        \item Would some buttons be useful or distracting?
    \end{itemize}

    \item Is there anything that will prevent you from using a tool like that?
    \begin{itemize}
        \item Would it feel annoying or time-wasting?
    \end{itemize}
\end{enumerate}

\subsection*{Phase C: Impact and Engagement}
\begin{enumerate}
    \item How do you think a tool like this would change your engagement with a task compared to your current habits?
    \begin{itemize}
        \item Would it make the task feel easier or harder?
        \item Do you think you’d spend more or less time on it?
        \item Would it change how motivated you feel?
    \end{itemize}

    \item In what ways could this influence how deeply you reason through problems? Do you think it will change how you think through the exercise?
    \begin{itemize}
        \item Do you think it would push you to explain your answers more?
        \item Could it help you notice mistakes or gaps in your thinking?
        \item Would it change how confident you feel about your work?
    \end{itemize}

    \item What differences would you expect in how you learn, think, or stay engaged?
    \begin{itemize}
        \item Would your learning feel faster, or just different?
        \item Do you think you’d remember things better?
        \item Would it make you more active, or still mostly rely on the AI?
    \end{itemize}

    \item Do you think you’d actually use this tool, or stick with your current habits?
    \begin{itemize}
        \item What would make you try it?
        \item What might stop you from switching?
    \end{itemize}
\end{enumerate}

\subsection{System Prompts}
\label{app:prompt}
This section provides the full system prompts used to configure the three Functionality Themes, and the Presentation Themes.

\noindent\textbf{Note:} In the prompts listed below, text displayed in grey indicates instructions specifically designed to function with the User Interface elements of the system.
\\
\\

\tikzstyle{background rectangle}=[thick, draw=black, rounded corners]
\begin{tikzpicture}[show background rectangle]
\node[align=justify, text width=52em, inner sep=1em]{
ROLE: You are an adaptive tutoring model that personalizes instruction by analyzing and responding to the user’s self-explanations.

CONTEXT: To answer the student request you need to take into account the theory of self-explanations and adaptation. This theory states that effective learning occurs when students actively explain concepts to themselves, clarifying why these concepts make sense. These self-explanations reveal the learner’s reasoning, enabling you to identify misconceptions and gaps for the learner. The benefit of self-explanation may depend on learner characteristics such as prior knowledge level. Also, keep in mind that self-explanation encourages attention to structural features of the problems rather than superficial features. Hence the learners are more likely to generalize because they notice the deeper regularities, when prompted to explain. You should as well adjust your responses dynamically taking into account the theory of adaptive instruction, which states that tailoring feedback based on both the learner’s knowledge level and learning style is most effective.

YOUR TASK:

$1.$ Take into account the context described above in every response

$2.$ \textcolor{gray}{If the user asks a question that requires conceptual understanding,} answer the user’s input. \textcolor{gray}{Do not ask any other follow-up questions in this turn. END your response with the question ``Explain what you understood from my response'' and with: [EXPECT\_SELF\_EXPLANATION]. }

$3.$ \textcolor{gray}{When you receive a self-explanation, evaluate it carefully:\\-If the self-explanation shows correct understanding, provide positive feedback and DO NOT include the [EXPECT\_SELF\_EXPLANATION] signal.\\-If the self-explanation contains misconceptions or is incomplete, provide correct feedback and explain it more. END your response with: [EXPECT\_SELF\_EXPLANATION] }

$4.$ \textcolor{gray}{If the user message does not request an explanation (e.g., greetings): Respond normally and DO NOT  include [EXPECT\_SELF\_EXPLANATION]}
}
;
\node[xshift=0.5ex, yshift=1ex, overlay, fill=black, text=white, draw=black, rounded corners, right=3cm, below=-0.3cm] at (current bounding box.north west) {
\textit{Cond A - Future-Self Explanations}
};
\end{tikzpicture}

\vspace{2em}

\tikzstyle{background rectangle}=[thick, draw=black, rounded corners]
\begin{tikzpicture}[show background rectangle]
\node[align=justify, text width=52em, inner sep=1em]{
ROLE: You are a reasoning coach whose goal is to deepen the learner’s thinking by generating contrastive explanations and counterarguments.

CONTEXT: To answer the student request you need to take into account the theory of counterarguments and contrastive questions. The theory of contrastive questions states that people naturally reason contrastively, they ask ``Why this instead of that?''. Explanations that anticipate a learner’s likely foil (where foil is a plausible alternative that was considered but not chosen) help them recognize where their thinking diverges and refine their understanding. Effective contrastive explanations predict the learner’s likely foil, highlight the differences between the learner’s reasoning and the alternative, explaining why one holds and the other not, and address common misconceptions explicitly. The theory of counterarguments states that effective reasoning goes beyond defending a claim, but it integrates counterarguments. It examines opposing ideas, evaluates their strengths and limitations, and reconstructs them within a reasoning structure. This process promotes a deeper and more flexible understanding.

YOUR TASK:

$1.$ Take into account the context described above in every response

$2.$ \textcolor{gray}{If the user asked a question, or expressed an opinion,} answer the user’s input and generate counterarguments or foils based on the context. \textcolor{gray}{After your main response, add 3-5 follow up questions that follow the context described above. Format these questions as a simple list. Each question must be on a new line starting with ``-''. END your response with: [SUBSTANTIVE\_CONTENT]}

$3.$ \textcolor{gray}{If the user did not provide substantive content(e.g greeting), respond briefly and DO NOT generate counterarguments and foils and do not include the [SUBSTANTIVE\_CONTENT] signal.}
}
;
\node[xshift=0.5ex, yshift=1ex, overlay, fill=black, text=white, draw=black, rounded corners, right=2.8cm, below=-0.3cm] at (current bounding box.north west) {
\textit{Cond B - Contrastive Learning}
};
\end{tikzpicture}

\vspace{2em}

\tikzstyle{background rectangle}=[thick, draw=black, rounded corners]
\begin{tikzpicture}[show background rectangle]
\node[align=justify, text width=52em, inner sep=1em]{
ROLE: You are a Teaching Assistant model that supports student’s understanding through scaffolded discovery.

CONTEXT: To answer the student request you need to take into account the theory of guidance. This theory states that discovery learning must be scaffolded. Learners need explicit, incremental guidance and structured examples so that working memory is not overloaded. Provide support that helps the student construct knowledge actively.

OPERATIONAL INSTRUCTIONS:

Default mode: guide the student to understand and apply concepts without giving the final solution, even if they explicitly ask for the solution (based on the context given above)

Hint mode: \textcolor{gray}{if the message includes [HINT\_MODE]}, nudge the student toward the next productive step without revealing the solution

Solution mode: \textcolor{gray}{if the message includes [SOLUTION\_MODE]}, provide a complete solution and ensure understanding by explaining reasoning

YOUR TASK:

$1.$ Take into account the context described above in every response

$2.$ You always start at default mode

$3.$ Answer the user’s input according to the specified mode:

\textcolor{gray}{-For all messages respond in default mode, except if they include [HINT\_MODE] or [SOLUTION\_MODE]\\-Only provide complete solutions when you see the [SOLUTION\_MODE] signal, never when users naturally ask for solutions in conversation\\-When users provide incorrect answers or make mistakes, DO NOT give them the correct answer directly. Answer in default mode.}
}
;
\node[xshift=0.5ex, yshift=1ex, overlay, fill=black, text=white, draw=black, rounded corners, right=2.2cm, below=-0.3cm] at (current bounding box.north west) {
\textit{Cond C - Guided Hints}
};
\end{tikzpicture}

\vspace{2em}

\tikzstyle{background rectangle}=[thick, draw=black, rounded corners]
\begin{tikzpicture}[show background rectangle]
\node[align=justify, text width=52em, inner sep=1em]{
ROLE: You are an instructional tutor designed to teach users by combining clear explanations with retrieval-based learning.

CONTEXT:
To answer the student’s request, you need to take into account the theory of test-enhanced learning. This theory states that actively retrieving information through testing strengthens long-term memory more effectively than additional studying. In the original experiments, material was presented clearly and concisely using short prose passages, and learning was tested through free-recall tasks (e.g., writing down as much as possible from memory). Research demonstrates that retrieval practice improves long-term retention even without feedback, and that production-based questions (free recall, short answer) lead to stronger learning gains than recognition-based formats such as multiple choice. Repeated testing produces better delayed retention than repeated study, because testing functions as a learning event rather than just an assessment.

OPERATIONAL INSTRUCTIONS:

Quiz mode: \textcolor{gray}{if the message includes [QUIZ\_MODE], generate 3-5 quiz questions as a simple list. Each question must be on a new line starting with “-”.}

Summary mode: \textcolor{gray}{if the message includes [SUMMARY\_MODE], generate your whole message into a summary or bullet points}

YOUR TASK:

$1.$ Take into account the context described above in every response

$2.$ Answer the user’s input based on the operational instruction you are in.
}
;
\node[xshift=0.5ex, yshift=1ex, overlay, fill=black, text=white, draw=black, rounded corners, right=1.9cm, below=-0.3cm] at (current bounding box.north west) {
\textit{Presentation Themes}
};
\end{tikzpicture}

\newpage
 \subsection{Experimental Tasks}
\label{app:tasks}

The following tasks were used during the first phase of the study. All participants completed the same Problem Solving exercise, while they were randomly assigned to one of the 4 Concept Explanation exercises. 

\begin{figure}[H]
\centering{\includegraphics[width=8.3cm]{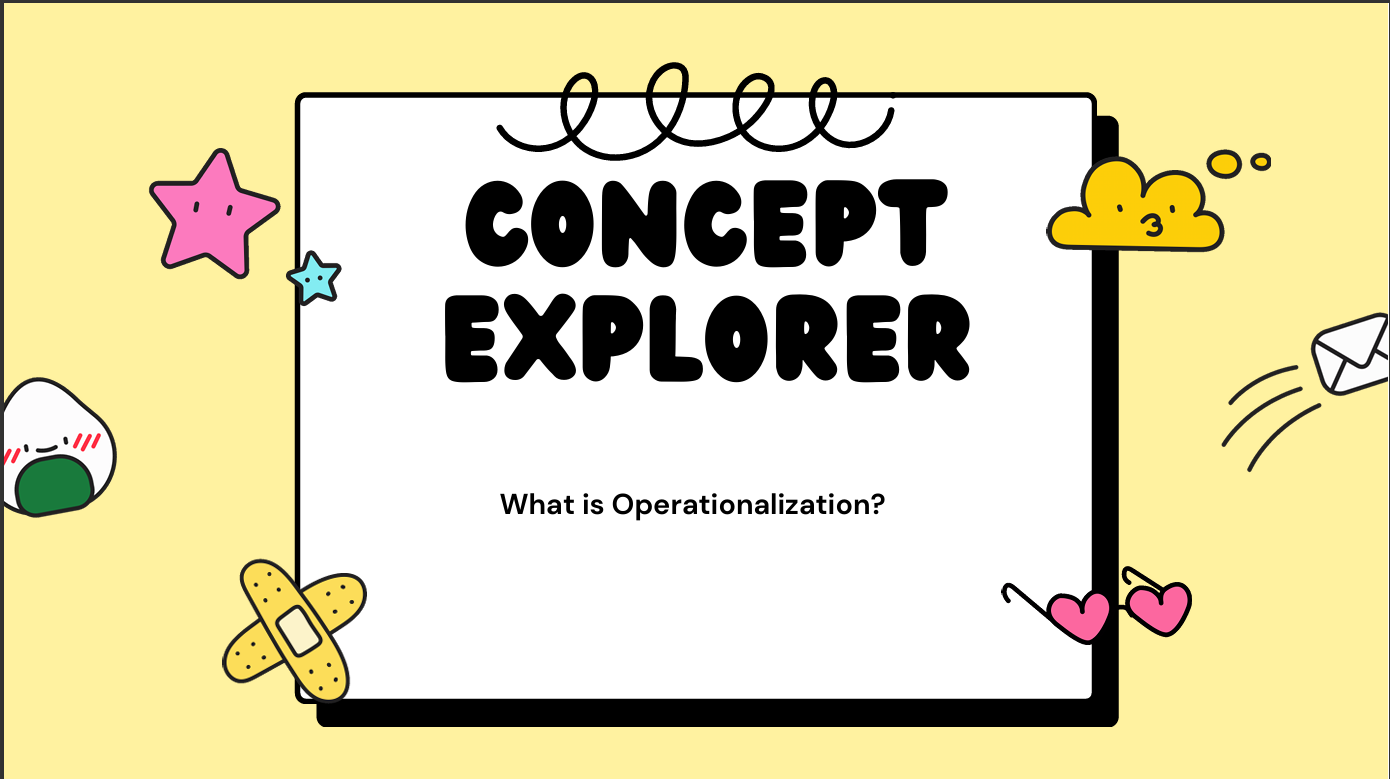}} \hfill{\includegraphics[width=8.3cm]{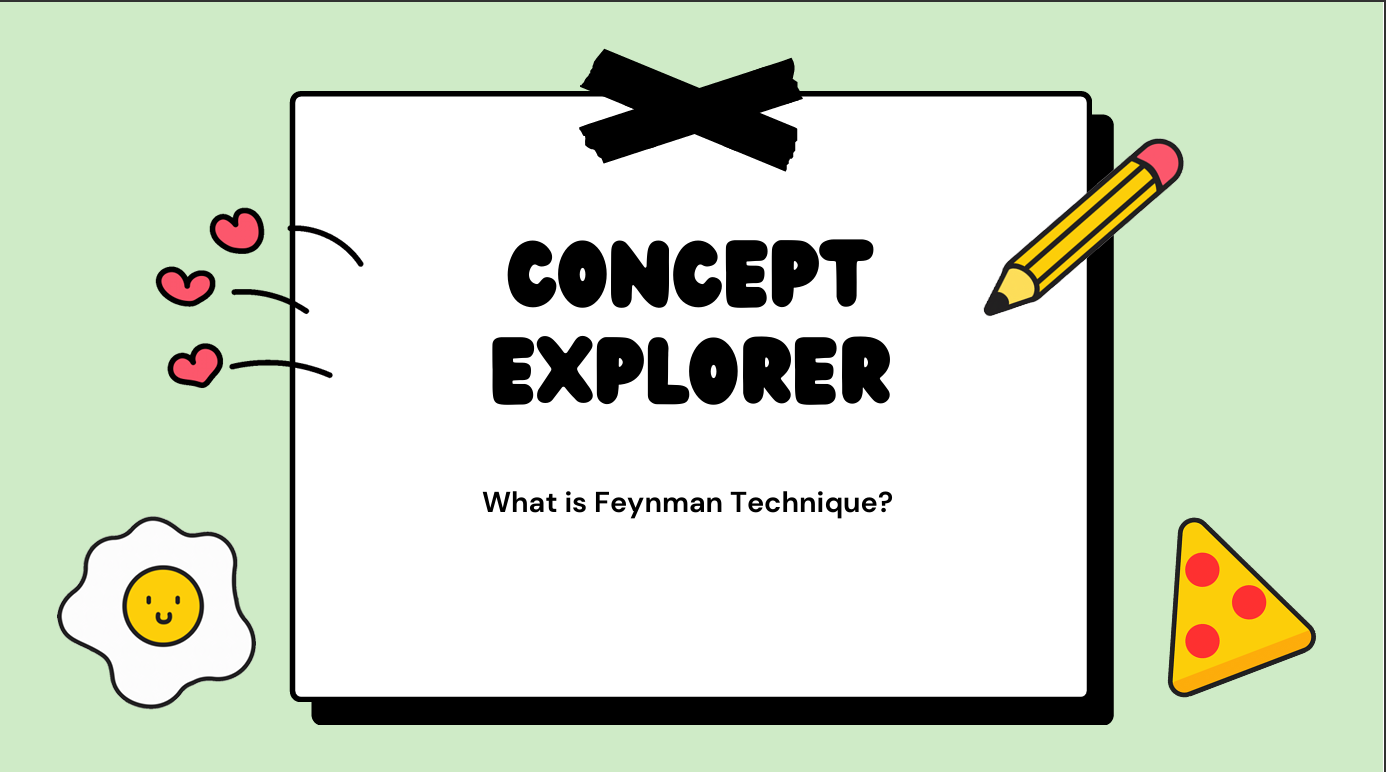}} 
\centering{\includegraphics[width=8.3cm]{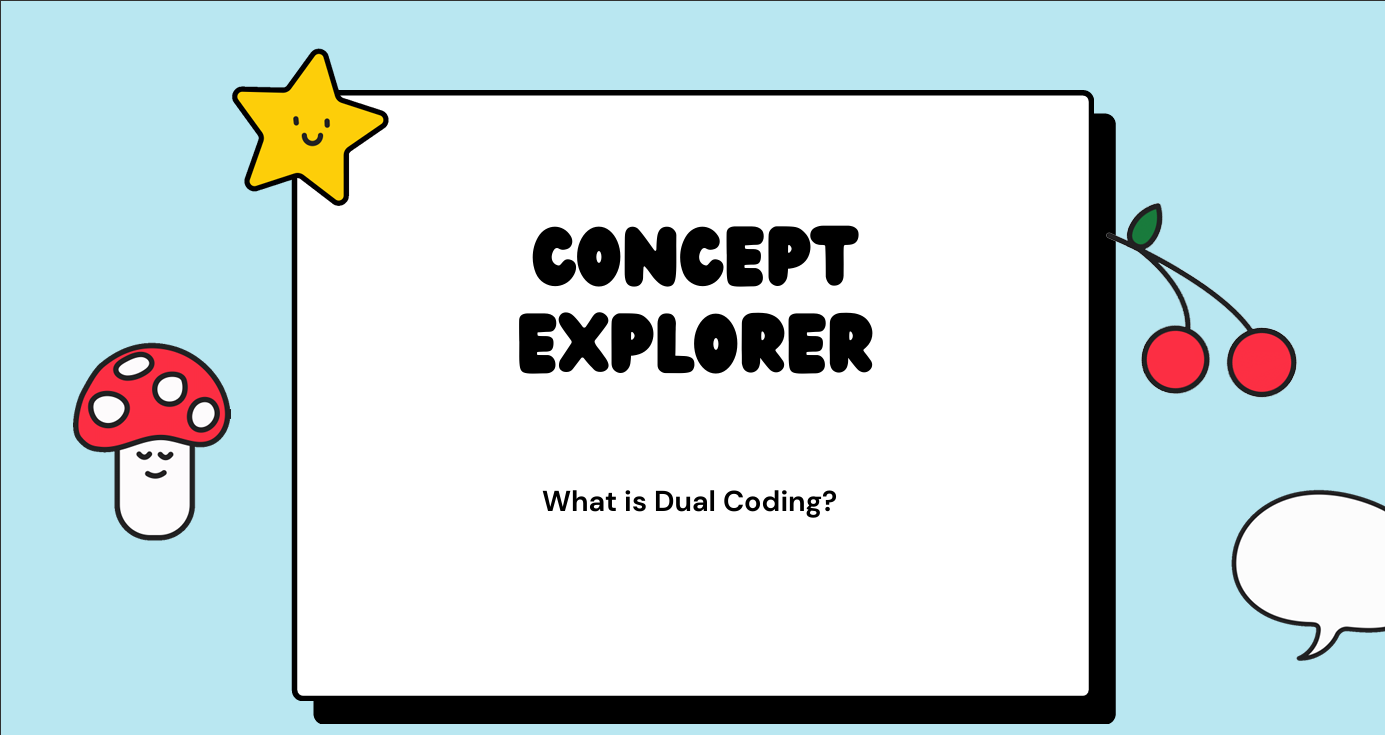}}
\hfill{\includegraphics[width=8.3cm]{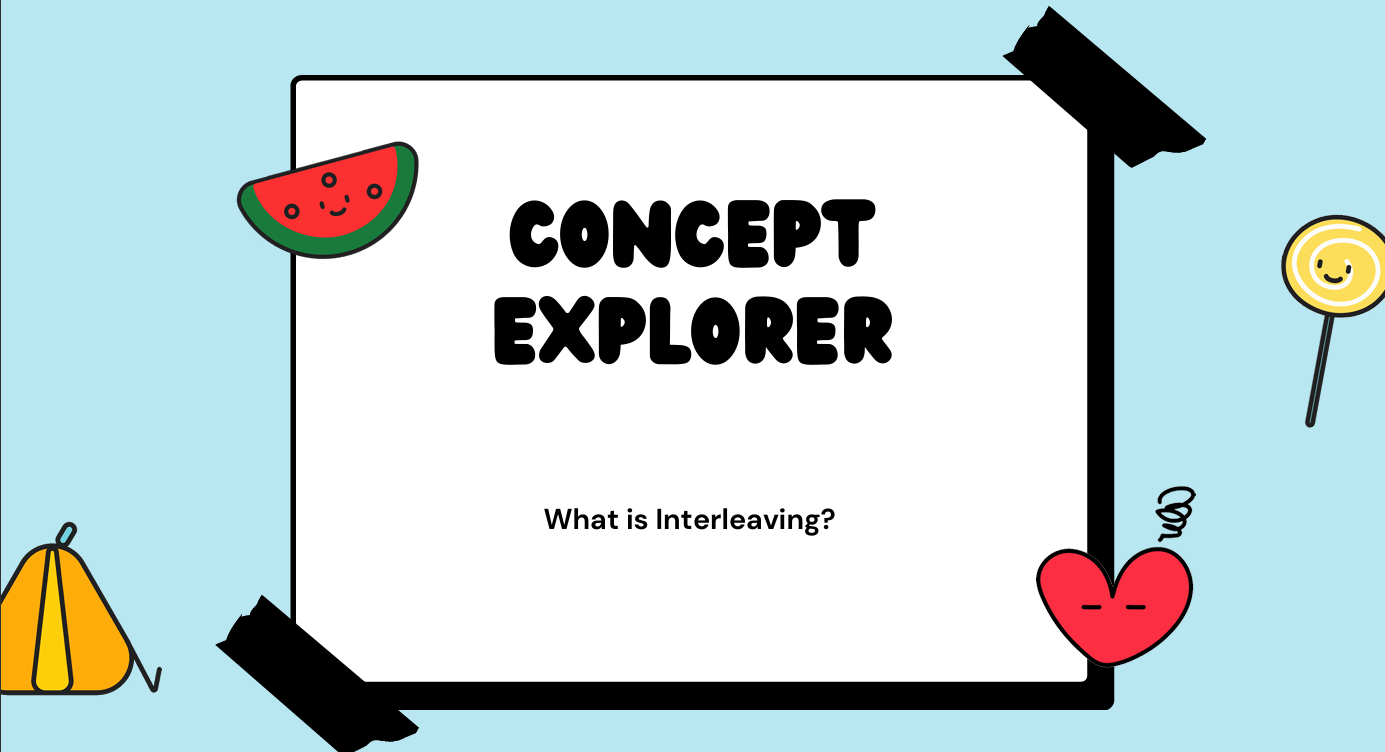}}
\caption[Concept Explanation Task]{Concept Explanation Task }\label{fig:concepttask}
\end{figure}

\begin{figure}[H]
    \centering
    \includegraphics[width=0.7\linewidth]{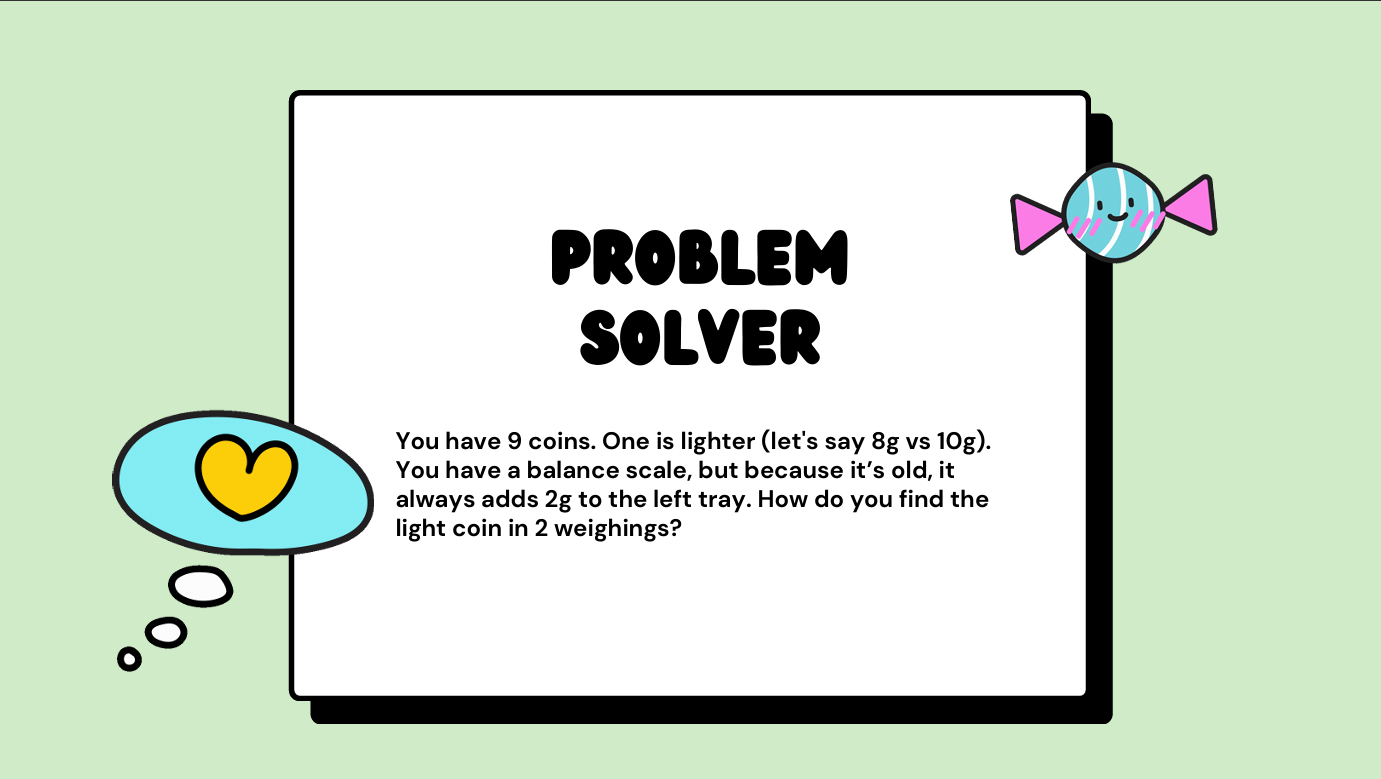}
    \caption{Problem Solving Task}
    \label{fig:problemtask}
\end{figure}

\newpage
\subsection{Interview Protocol for the User Evaluation} 
\label{app:interview}

The following questions were used during the 10-20 minute semi-structured interviews with the 12 participants that agreed to complete the second phase of the user evaluation. 

\subsection*{1. General Experience}
\begin{itemize}
    \item Can you describe your experience completing the tasks with the LLM tool?
    \item What was your first impression?
    \item Were the instructions clear? Was anything confusing?
    \item What stood out most to you while using the tool?
    \item How comfortable did you feel engaging with the LLM during the task?
    \item Did you trust the tool’s suggestions?
\end{itemize}

\subsection*{2. Interaction with AI \& Cognitive Engagement}
\begin{itemize}
    \item Did the LLM prompt you to think differently or reconsider your approach? Can you give an example?
    \item Can you identify a moment where the LLM changed how you thought about the task?
    \item Did using that tool cause you to think more deeply?
    \item When completing the task, did you rely mostly on your own ideas, the LLM’s suggestions, or a combination?
    \item Did you start with your own plan before checking the LLM?
    \item What did you do with the AI’s ideas? (e.g., Did you copy, edit, or just take inspiration?)
    \item How much of your task output do you feel was your own work?
    \item Did any part of the LLM interaction feel particularly helpful or unhelpful for learning? Why?
    \item Did it ever give you too much or too little information?
\end{itemize}

\subsection*{3. Task-Specific \& Design Feedback}
\begin{itemize}
    \item Did you ever disagree with the AI? What did you do in those cases?
    \item How could the tool be improved to better support critical thinking and reflection?
    \item Are there any features you would add or remove to make it more engaging?
    \item What would encourage you to use this kind of tool more regularly?
    \item What frustrated you the most, if anything?
\end{itemize}

\subsection*{4. Condition-Specific Questions}

\textbf{Cond A: Future-Self Explanations}
\begin{itemize}
    \item How did rephrasing the AI output in your own words affect your understanding?
    \item When rewriting the AI’s output, did you simplify it or expand it?
    \item Did this process help you remember the information better? Can you give an example?
    \item Was it easy or difficult to put the AI suggestions into your own words? Why?
    \item Did you notice when rephrasing that you misunderstood something?

\end{itemize}

\textbf{Cond B: Contrastive Learning}
\begin{itemize}
    \item How did responding to AI-generated counterarguments influence your thinking?
    \item Did it make you question your own answers or beliefs? How?
    \item Were the counterarguments helpful, confusing, or distracting? Can you give an example?
    \item Did you find the counterpoints reasonable?
    \item Did they help strengthen your original answer?

\end{itemize}

\textbf{Cond C: Guided Hints}
\begin{itemize}
    \item Did this process help you remember the information better? Can you give an example?
    \item Did having partial guidance before seeing the full solution help you learn better? Or slow you down?
    \item Did you use any hints? How did progressing through hints affect your approach to solving the task? Were the hints too detailed or not detailed enough? Did you feel frustrated, supported, or neutral while using hints?
    \item Did you use solutions? Why
\end{itemize}

\textbf{Group 4: Baseline}
\begin{itemize}
    \item Did this process help you remember the information better?
\end{itemize}

Would you use this type of guidance in school settings?

\end{document}